\documentclass[aps,prd,twocolumn,nofootinbib]{revtex4-1}

\usepackage{graphicx}
\usepackage{slashed}

\begin{document}

\title{$B_c$ Meson Production around the $Z^0$ Peak at a High Luminosity $e^+ e^-$ Collider}

\author{Zhi Yang$^1$}
\author{Xing-Gang Wu$^{1,2}$}
\email{wuxg@cqu.edu.cn}
\author{Gu Chen$^{1}$}
\author{Qi-Li Liao$^{1}$}
\author{Jia-Wei Zhang$^3$}

\address{$^{1}$ Department of Physics, Chongqing University, Chongqing 401331, P.R. China\\
$^2$ SLAC National Accelerator Laboratory, 2575 Sand Hill Road, Menlo Park, CA 94025, USA\\
$^3$ Department of Physics, Chongqing University of Science and Technology, Chongqing 401331, P.R. China}

\date{\today}

\begin{abstract}
Considering the possibility to build an $e^+ e^-$ collider at the energies around the $Z^0$-boson resonance with a planned luminosity so high as ${\cal L}\propto 10^{34}\sim 10^{36}cm^{-2}s^{-1}$ (super $Z$-factory), we make a detailed discussion on the $(c\bar{b})$-quarkonium production through $e^{+}+e^{-}\rightarrow (c\bar{b})[n]+b+\bar{c}$ within the framework of non-relativistic QCD. Here $[n]$ stands for the Fock-states $|(c\bar{b})_{\bf 1}[^1S_0]\rangle$, $|(c\bar{b})_{\bf 8}[^1S_0]g\rangle$, $|(c\bar{b})_{\bf 1}[^3S_1]\rangle$, $|(c\bar{b})_{\bf 8}[^3S_1]g\rangle$, $|(c\bar{b})_{\bf 1}[^1P_1]\rangle$ and $|(c\bar{b})_{\bf 1}[^3P_J]\rangle$ (with $J=(1,2,3)$) respectively. To simplify the hard-scattering amplitude as much as possible and to derive analytic expressions for the purpose of future events simulation, we adopt the ``improved trace technology" to do our calculation, which deals with the hard scattering amplitude directly at the amplitude level other than the conventional way at the squared-amplitude level. Total cross-section uncertainties caused by the quark masses are predicted by taking $m_c=1.50\pm0.30$ GeV and $m_b=4.90\pm0.40$ GeV. If all higher $(c\bar{b})$-quarkonium states decay to the ground state $B_c$ ($|(c\bar{b})_{\bf 1}[^1S_0]\rangle$) with $100\%$ efficiency, we obtain $\sigma_{e^{+}+e^{-}\rightarrow B_{c}+b+\bar{c}} =5.190^{+6.222}_{-2.419}$ pb, which shows that about $10^5 \sim 10^7$ $B_c$ events per operation year can be accumulated in the super $Z$-factory. If taking the collider energy runs slightly off the $Z^0$-peak, i.e. $\sqrt{S}=(1.00\pm0.05) m_Z$, the total cross-section shall be lowered by about one-order from its peak value. Such a super $Z$-factory shall provide another useful platform to study the properties of $B_c$ meson, or even the properties of its excited $P$-wave states, in addition to its production at the hadronic colliders Tevatron and LHC.  \\


\end{abstract}

\maketitle

\section{Introduction}

Recently, a Chinese group has raised the proposal of constructing a high luminosity $e^+ e^-$ collider in China, the so-called super $Z$-factory \cite{wjw}, which is similar to the Gigaz program suggested by the Internal Linear Collider Collaboration \cite{ilc,gigaz} but with a even higher luminosity. As for this newly purposed super $Z$-factory, a $e^+ e^-$ collider shall run at the energies around the mass of $Z^0$-boson resonance and with a high luminosity up to ${\cal L}\propto 10^{34}\sim 10^{36}cm^{-2}s^{-1}$. Many interesting topics can be studied in this super $Z$-factory: the properties of $Z^0$-boson and the precision tests of the electro-weak sector of the standard model, the properties of the heavy mesons or baryons, $\tau$ physics, the physics beyond the standard model and etc.

The $B_c$ meson has aroused people's great interests since its discovery by CDF collaboration \cite{cdf1,cdf2}. The $B_c$ meson is a doubly heavy quark-antiquark bound state and carries flavors explicitly. Systematic studies of its production properties at the hadronic colliders Tevatron and LHC have been done in the literature, e.g. Refs.\cite{bc1,bc2,bc3,tbc,tbc0,wbc1,wbc2}. In particular, a computer program BCVEGPY for the hadronic production of the $B_c$ meson has been completed and developed in Refs.\cite{bcvegpy1,bcvegpy2}, which can be conveniently implemented into PYTHIA \cite{pythia} for simulating $B_c$ events. Very recently, a first observation of the decay channel $B_c^+ \to J/\Psi \pi^+ \pi^- \pi^{+}$ with the help of BCVEGPY by LHCb group has been finished \cite{lhcbnew}. At the hadronic colliders, there is much pollution from the hadronic background and many produced $B_c$ events shall be cut off by the trigging condition \cite{yellow,cms}. So, some alternative measurements would be helpful for a comprehensive study.

Comparing to the hadronic colliders as the Tevatron and LHC, a $e^+e^-$ collider has three general features as cleanliness, democracy and holism \cite{peskin}, which are helpful and have some advantages to perform more precise measurements for certain processes. To seek the $B_c$ events at a $e^+ e^-$ collider has already been tried by the LEP-I experiment, however no $B_c$ events have been found there. If the luminosity of a $e^+e^-$ collider is raised up to ${\cal L}\propto 10^{34-36}cm^{-2}s^{-1}$ and its colliding energy is confined to be around the $Z^0$-peak, it will raise the production rate up to several orders. This is reasonable, since as estimated in Ref.\cite{gigaz}, more than $10^{9\sim11}$ $Z^0$-boson resonance can be produced for Gigaz with ${\cal L}\propto 10^{34}cm^{-2}s^{-1}$. Moreover, Refs.\cite{zbc0,bchad2,zbc1,zbc2,zbc3,zbc4} have shown that sizable $B_c$ events can be produce through the $Z^0$-boson decays. Then, as a rough combination, one can estimate that such a super $Z$-factory will open new opportunities for studying the $B_c$-meson properties. And to be useful reference, it is interesting to present a more sound estimation.

Under the non-relativistic QCD (NRQCD) framework \cite{nrqcd}, a doubly heavy meson is considered as an expansion of Fock states, and the probability of finding each Fock state in the expansion is proportional to a definite power of $v$, where $v$ is the relative velocity and $v^2$ is around $0.1\sim0.3$ for heavy quarkonium. We have studied the production of $(c\bar{c})$- and $(b\bar{b})$- quarkonium with a photon or via initial state radiation at the $Z^0$-peak in the super $Z$-factory \cite{cwwbc}. It is interesting to see whether there has enough $B_c$ events produced at such a high luminosity $e^+ e^-$ collider, which can be produced through the process $e^{+}+e^{-}\rightarrow \gamma^*/Z^0 \rightarrow (c\bar{b})[n]+b+\bar{c}$, where $[n]$ stands for the $(c\bar{b})$-quarkonium Fock states. Excited $(c\bar{b})$-quarkonium states, such as the color-singlet $P$-wave states and the color-octet $S$-wave states, may directly or indirectly decay to the ground state with almost 100\% possibility via electromagnetic or hadronic interactions. Then, as a full estimation of the $B_c$-meson production, we shall discuss the following $(c\bar{b})$-quarkonium states simultaneously; i.e. $|(c\bar{b})_{\bf 1}[^1S_0]\rangle$, $|(c\bar{b})_{\bf 8}[^1S_0]g\rangle$, $|(c\bar{b})_{\bf 1}[^3S_1]\rangle$, $|(c\bar{b})_{\bf 8}[^3S_1]g\rangle$, $|(c\bar{b})_{\bf 1}[^1P_1]\rangle$ and $|(c\bar{b})_{\bf 1}[^3P_J]\rangle$ (with $J=(1,2,3)$). Here the thickened subscript of $(c\bar{b})$-quarkonium denotes the color index, {\bf 1} for color-singlet and {\bf 8} for color-octet; the relevant angular momentum quantum numbers are shown in the parentheses accordingly.

Because of the emergence of massive-fermion lines, the analytical expression for the squared amplitude becomes too complex and lengthy for more (massive) particles in the final states and for more higher-order Fock states to be generated. For example, to derive the amplitudes for the $P$-wave states, one also needs to get the derivative of the amplitudes over the relative momentum of the constitute quarks. It has been found that to do the numerical calculation using the conventional squared-amplitude technology becomes time-consuming for these complex processes, since the cross-terms increases with the increment of Feynman diagrams, $|{\cal M}|^2=\sum_{ij}{\cal M}_{i} {\cal M}^*_{j}$, where $i$ stands for the number of Feynman diagrams for the process. One important way to solve this is to deal with the process directly at the amplitude level. For the purpose, the ``improved trace technology" is suggested and developed in the literature \cite{itt0,wbc2,zbc0,zbc1,zbc2}. After generating proper phase-space points, one first calculate the numerical value for the amplitudes, and then sum these values algebraically and square it to get the squared amplitude, $|{\cal M}|^2=|\sum_{i}{\cal M}_{i}|^2$; through such way, numerical simulation efficiency can be greatly improved in comparison to the usual squared amplitude technology \footnote{One can also generate $B_c$ events through PYTHIA by $e^+e^-$ annihilation. However, in PYTHIA all kinds of $b$-hadron ( $B_s$, $B_c$, $\Lambda_b$ and etc. ) events will be produced according to the proper fragmentation possibilities, among which the fragmentation possibility for the doubly heavy $B_c$ meson is quite small \cite{zbc4}, thus using PYTHIA to generate $B_c$ events is not `economical'; i.e., when simulating the $B_c$ events, too much time will be wasted by generating enormous unwanted $b$-hadron events. }. Moreover, under the approach, many simplifications can be done at the amplitude level due to the fermion-line symmetries and the specific properties of each heavy-quarkonium Fock states, then, we can even written down the analytic expressions for the amplitude.

The ``Improved trace technology" has been adopted to derive a simplified and compact analytical expressions for the on-shell $Z^0$-boson decay process, $Z^0 \rightarrow (c\bar{b})[n]+b+\bar{c}$ \cite{zbc0,zbc1,zbc2}. At the present, we will adopt the same method for dealing with the $(c\bar{b})$-quarkonium production through the full production process, $e^{+}+e^{-}\rightarrow \gamma^*/Z^0 \rightarrow (c\bar{b})[n]+b+\bar{c}$. The intermediate $Z^0$-boson may or may not be on-shell, and if it is off-shell, some new Lorentz structures must be introduced to ensure a gauge-invariant result. In a sense, this paper provides an extension of the ``improved trace technology" on how to deal with the off-shell particles.

The remaining parts of the paper are organized as follows. In Sec.II, we present the formulation for dealing with the process of $e^{+}+e^{-}\rightarrow Z^0/\gamma^* \rightarrow (c\bar{b})[n]+b+\bar{c}$, where improved trace technology are adopted to simplify the hard scattering amplitude. In Sec.III, we give the numerical results. Sec.IV is reserved for a summary. All necessary expressions are put in the Appendices.

\section{Formulation and Calculation techniques}

\begin{figure}
\includegraphics[width=0.40\textwidth]{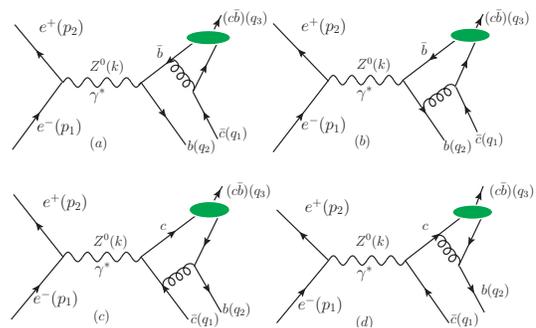}
\caption{Feynman diagrams for the process $e^{+}+e^{-}\rightarrow \gamma^{*}/Z^0\rightarrow (c\bar{b})[n]+b+\bar{c}$, where $[n]$ stands for the Fock states: $|(c\bar{b})_{\bf 1}[^1S_0]\rangle$, $|(c\bar{b})_{\bf 8}[^1S_0]g\rangle$, $|(c\bar{b})_{\bf 1}[^3S_1]\rangle$, $|(c\bar{b})_{\bf 8}[^3S_1]g\rangle$, $|(c\bar{b})_{\bf 1}[^1P_1]\rangle$ and $|(c\bar{b})_{\bf 1}[^3P_J]\rangle$ (with $J=(1,2,3)$) respectively. } \label{feyn}
\end{figure}

As shown in Fig.(\ref{feyn}), there are four Feynman diagrams for the process $e^{+}(p_2) +e^{-}(p_1)\rightarrow Z^0/\gamma^* \rightarrow (c\bar{b})[n] (q_3)+b(q_2)+\bar{c}(q_1)$, where $[n]$ stands for the $(c\bar{b})$-quarkonium Fock-states: $|(c\bar{b})_{\bf 1}[^1S_0]\rangle$, $|(c\bar{b})_{\bf 8}[^1S_0]g\rangle$, $|(c\bar{b})_{\bf 1}[^3S_1]\rangle$, $|(c\bar{b})_{\bf 8}[^3S_1]g\rangle$, $|(c\bar{b})_{\bf 1}[^1P_1]\rangle$ and $|(c\bar{b})_{\bf 1}[^3P_J]\rangle$ (with $J=(1,2,3)$) respectively. Here as a comparison, the production channel via a virtual photon, which is small around the $Z^0$ peak, is also presented.

According to NRQCD factorization formulae \cite{petrelli}, the cross-section of the process can be written in the following factorization form:
\begin{equation}
d\sigma = \sum_{n} d\hat\sigma (e^{+}+e^{-}\rightarrow (c\bar{b})[n]+b+\bar{c})\langle{\cal O}^H(n)\rangle,
\end{equation}
where the matrix element $\langle{\cal O}^H(n)\rangle$ is proportional to the inclusive transition probability of the perturbative state $(c\bar{b})[n]$ into the bound states.  $d\hat\sigma(e^{+}+e^{-}\rightarrow (c\bar{b})[n]+b+\bar{c})$ stands for the short-distance cross-section; i.e.
\begin{eqnarray}
&&d\hat\sigma(e^{+}+e^{-}\rightarrow (c\bar{b})[n]+b+\bar{c}) \nonumber\\
&& \quad = \frac{1}{4\sqrt{(p_1\cdot p_2)^2-m_1^2 m_2^2}} \overline{\sum}  |{\cal M}|^{2} d\Phi_3,
\end{eqnarray}
where $\overline{\sum}$ means we need to average over the spin states of initial particles and to sum over the color and spin of all final particles, and the three-particle phase space
\begin{displaymath}
d{\Phi_3}=(2\pi)^4 \delta^{4}\left(p_1+p_2 - \sum_f^3 q_{f}\right)\prod_{f=1}^3
\frac{d^3{q_f}}{(2\pi)^3 2q_f^0}.
\end{displaymath}
The phase space can be generated by the routine RAMBOS \cite{rambos}, and its integration can be carried out numerically with the VEGAS program \cite{vegas}.

The scattering amplitude for the process can be written as,
\begin{equation}
i{\cal M}={\cal C}\times L_{rr^{\prime}}^{\mu}D_{\mu\nu}H_{ss^{\prime}}^{\nu} ,
\label{A1}
\end{equation}
where
\begin{equation}
L_{rr^{\prime}}^{\mu} = \bar{v}_r(p_2) \Gamma^{\mu} u_{r^{\prime}}(p_1)
\end{equation}
and
\begin{equation}
H_{ss^{\prime}}^{\nu} = \bar{u}_s(q_2)\sum\limits_{n=1}^{4}{\cal A}^{\nu}_n v_{s^{\prime}}(q_1) .
\end{equation}
The spin-indices $r$, $r^{\prime}$, $s$ and $s^{\prime}$ stand for the spin projections of the initial electron and positron, and the outgoing quark and antiquark respectively. The overall parameter ${\cal C}={\cal C}_s$ or ${\cal C}_o$ is for the production of the color-singlet or the color-octet quarkonium state respectively. For the quarkonium production through $Z^0$-boson propagator: ${\cal C}_s=\frac{4}{3\sqrt{3}}\frac{e^2 g^2_s}{\sin^2\theta_w\cdot \cos^2\theta_w}\delta_{ij}$ and ${\cal C}_o=\frac{e^2 g_s^2}{\sin^{2}\theta_{w} \cos^{2}\theta_w}\times (\sqrt{2}T^aT^bT^a)_{ij}$ ($\sqrt{2}T^b$ stands for the color of the color-octet $(c\bar{b})$-quarkonium state); the vertex $\Gamma^{\mu}=\gamma^\mu(1-4\sin^2\theta_w-\gamma^5)$ and the propagator $D_{\mu\nu}=\frac{i}{k^2-m^2_Z +im_Z\Gamma_z}\left(-g_{\mu\nu}+{k_\mu k_\nu}/{k^2}\right)$, where $\Gamma_z$ stands for the total decay width of $Z^0$ boson. If the $e^+ e^-$ collider runs at the energies around the $Z^0$ mass, it will constitute an energy scan. The shape of the cross-section variation around the $Z^0$ peak is described by a Breit-Wigner ansatz with an energy-dependent total width. While, for the quarkonium production through photon propagator: ${\cal C}_s=\frac{4}{3\sqrt{3}}{e^2 g^2_s}\delta_{ij}$ and ${\cal C}_o={e^2 g_s^2}(\sqrt{2}T^aT^bT^a)_{ij}$; the vertex $\Gamma^{\mu}=\gamma^\mu$ and the propagator $D_{\mu\nu}=\frac{-i}{k^2}g_{\mu\nu}$. The gamma structures ${\cal A}^{\nu}_n$ with $n=1,\cdots,4$ can be read off from the four Feynman diagrams in Fig.(\ref{feyn}), whose explicit expressions can be directly found in Refs.\cite{zbc1,zbc2} \footnote{As for the case of $Z^0$ propagator, from the correspondence between ${\cal A}^{\nu}_n$ and ${\cal A}_n$, one only needs to replace $\slashed{\epsilon}(k)$ in the amplitudes of Refs.\cite{zbc1,zbc2} by ${\gamma^\nu}$. And the case of photon propagator can be easily derived by changing the corresponding vertex.}.

$L_{rr^{\prime}}^{\mu}$ and $H_{ss^{\prime}}^{\nu}$ contain explicit strings of Dirac $\gamma$-matrices sandwiched by two fermion spinors, which can be done under the conventional trace technology. However it is hard to write down the squared amplitudes for these massive processes, especially for the $P$-wave cases, which is too complex and lengthy. Here, to simplify the analytic results and to make it more compact, we adopt the ``improved trace technology" \cite{wbc2,zbc0,zbc1,zbc2}. It is noted that under suitable transformations, this approach is also useful for deriving the polarized cross-sections, which is helpful for a polarized $e^+ e^-$ collider.

As an explanation for the ``improved trace technology", we first arrange the amplitude $H_{ss^{\prime}}^{\nu}$ into four orthogonal sub-amplitudes $H_{i}^{\nu}$ according to the spins of the outgoing quark $q$ with spin $s$ and antiquark $\bar{q}$ with spin $s'$, then transform these sub-amplitudes into a trace form by properly dealing with the massive spinors with the help of an arbitrary light-like momentum $k_0$ and an arbitrary space-like momentum $k_1$, which satisfies $k_1^2=-1$ and $k_0\cdot k_1 =0$. The final results are independent of $k_0$ and $k_1$, which provides a way to check the rightness of the derived results. And one can choose them to be those that will maximumly simply the amplitude. And then we do the trace of the Dirac $\gamma$-matrix strings at the amplitude level, which shall result in explicit series over some independent Lorentz-structures. The amplitude $L_{rr^{\prime}}^{\mu}$ can be done in a similar way. Detailed process of the approach can be found in Refs.\cite{zbc1,zbc2}, so to short the paper, we only list all the differences for the present processes. However, because at the present $Z^0$ may or may not be on-shell, one should use these on-shell expressions carefully with suitable transformations.

As for $H_{ss^{\prime}}^{\nu}$, it can be rearranged as the following four spin combinations
\begin{eqnarray}
H^{\nu}_1 &=& \frac{H^{\nu}_{ss'}+H^{\nu}_{-s-s'}}{\sqrt{2}} \\
H^{\nu}_2 &=& \frac{H^{\nu}_{ss'}-H^{\nu}_{-s-s'}}{\sqrt{2}} \\
H^{\nu}_3 &=& \frac{H^{\nu}_{s-s'}-H^{\nu}_{-ss'}}{\sqrt{2}} \\
H^{\nu}_4 &=& \frac{H^{\nu}_{s-s'}+H^{\nu}_{-ss'}}{\sqrt{2}} .
\end{eqnarray}
The production through the photon propagator is similar to the case of the $Z^0$-boson propagator, which can be directly read from its vector part. So we shall only concentrate our attention on the case of the $Z^0$-boson propagator \footnote{A generator BEEC for dealing with the heavy quarkonium production at a $e^+ e^-$ collider within the PYTHIA environment \cite{pythia} is in preparation, where both cases with the photon and the $Z^0$-boson propagators are included \cite{beec}.}. $H_{1,2,3,4}^{\nu}$ can be expressed over the independent Lorentz structures. For example, for the production of $|(c\bar{b})_{\bf 1}[^1S_0]\rangle$-quarkonium, we have
\begin{eqnarray}
H^{\nu}_{[^1S_0]} &=& {A_{1}^{[^1S_0]}}\frac{q_3^{\nu}}{m_Z}+{A_{2}^{[^1S_0]}}\frac{q_2^{\nu}}{m_Z}+ {A_{3}^{[^1S_0]}}\frac{k^{\nu}}{m_Z} \nonumber\\
&& +{A_{4}^{[^1S_0]}}\frac{i}{m_Z^3}\varepsilon(\nu,k,q_2,q_3) , \label{hmat}
\end{eqnarray}
where $\varepsilon(\nu,k,q_2,q_3)=\varepsilon^{\nu\beta\rho\sigma} k_{\beta} q_{2\rho} q_{3\sigma}$. Comparing the Lorentz expansions of $H^{\nu}_{[^1S_0]}$ with those of $M_{[^1S_0]}$ in Refs.\cite{zbc1,zbc2}, we obtain the same expansion coefficients for the corresponding Lorentz structures. More explicitly, by multiplying both side of Eq.(\ref{hmat}) with the on-shell $Z^0$-boson polarization vector $\epsilon_{\nu}(k)$, which satisfies $k\cdot\epsilon(k)=0$, we can obtain exactly the same matrix elements i.e. $\epsilon_{\nu}(k)H^{\nu}_{{[^1S_0]}}|_{k\cdot\epsilon=0}=M_{[^1S_0]}$. However, when the $Z^0$-boson is not on-shell, $\left({A_{3}^{[^1S_0]}} {k^{\nu}}/{m_Z}\right)$ must be kept in $H^{\nu}_{[^1S_0]}$. And to be a consistent analysis for the present case, we need to take these new terms into consideration. For convenience, we present all the new Lorentz structures for all the mentioned Fock states together with their non-zero coefficients in Appendix B.

As for the leptonic vector $L_{rr^{\prime}}^{\mu}$, it is simpler and can be directly expressed as
\begin{eqnarray}
L^{\mu}_{rr^{\prime}} &=& {C^{rr^{\prime}}_1}\frac{p_1^{\mu}}{m_Z}+
{C^{rr^{\prime}}_2}\frac{p_2^{\mu}}{m_Z}+{C^{rr^{\prime}}_3}\frac{q_3^{\mu}}{m_Z} \nonumber\\
&& +{C^{rr^{\prime}}_4}\frac{i}{m_Z^3}\varepsilon(\mu,p_1,p_2,q_3),
\end{eqnarray}
where all the non-zero coefficients $C^{rr^{\prime}}_{n}$ are presented in Appendix C.

To be a cross check, we also use the squared amplitude approach to deal with the process, which however can be rewritten as the following more convenient form
\begin{equation}
|{\cal M}|^2=\frac{1}{3}{\cal C}^2\times(D_{\alpha\beta}L^{\alpha\beta}) \cdot (D^{\rho\sigma}H_{\rho\sigma}).
\end{equation}
The leptonic tensor
\begin{equation}
L^{\alpha\beta}=-4 g^{\alpha\beta} {p_1}\cdot {p_2}+4 {p_1}^{\beta} {p_2}^{\alpha}+4 {p_1}^{\alpha}{p_2}^{\beta} ,
\end{equation}
where small mass terms involving $m_e$ are neglected. The result for the phase-space integration over $(D^{\rho\sigma}H_{\rho\sigma})$ is equal to the decay width of $Z^0\rightarrow(c\bar{b})[n]+b+\bar{c}$ times a factor ${1}/{\left[(k^2-m^2_Z)^2+ m^2_Z\Gamma^2_Z\right]}$. Numerically, we obtain the same results for these two approaches.

\section{Numerical results}

The basic input parameters are chosen as the following values: \cite{pdg}: $m_Z=91.1876$ GeV, $\Gamma_z=2.4952$ GeV, ${\sin ^2}{\theta _w}=0.23119$ and $\alpha_s(m_Z)=0.1176$. We adopt the leading-order $\alpha_s$ running, and take the normalization scale to be $2m_c$, which leads to $\alpha_s(2m_c)=0.212$. The color-singlet matrix elements can be directly related to the wave functions at the origin for the $S$-wave states or with the first derivative of the wave functions at the origin for the $P$-wave states. The $S$-wave color-octet ones are suppressed to the corresponding ones by a factor $\Delta^2_S(v)$ with $\Delta_S(v)\sim v^2$ \cite{yellow,nrqcd,bcdecay}. In the leading order approximation, the color-octet or color-singlet matrix-element appears in the amplitude as a linear factor. The uncertainties in the production from the matrix elements can be figured out straightforwardly, so throughout the paper, we will fix the values of the radial wave function at the origin and the first derivative of the radial wave function at the origin to be $|R_S(0)|^2=1.642$ GeV$^3$ and $|R_P^{\prime}(0)|^2=0.201$ GeV$^5$ \cite{wforigin} which are derived under the Buchm\"{u}ller-Tye-potential model \cite{wb}.

To calculate the hard-scattering amplitude for the short-distance cross-section $d\hat\sigma$, the equations: $q_{c}^2=m_c^2$, $q_{\bar b}^2=m_b^2$ and $q^2_{3}=M^2$  ($M$ is the $(c\bar{b})$-quarkonium mass, $q_{3}=q_{c}+q_{\bar b}$ with $q_c$ and $q_{\bar{b}}$ are momenta of the constitute quarks) must be satisfied simultaneously, and only when these equations are satisfied can the gauge invariance of the amplitude be guaranteed. In this paper, we proceed the numerical calculation by taking $m_c=1.50$ GeV and $m_b=4.90$ GeV as their center values \footnote{There values are consistent with the so called 1S masses \cite{1Smass}, i.e. half of the $J/\Psi$-mass for charm quark and half of the $\Upsilon(1S)$ for bottom quark. And they are also consistent with the bound-state parameters under the potential-model calculations \cite{wforigin,wang}. }. The mass of the quarkonium bound-state is the sum of the two heavy quark masses; i.e. $M=m_{b}+m_{c}=6.40$ GeV. We will discuss the theoretical uncertainties caused by the $c$- and $b$- quark masses by varying $m_c=1.50\pm0.30$ GeV and $m_b=4.90\pm0.40$ GeV.

\begin{figure}
\includegraphics[width=0.35\textwidth]{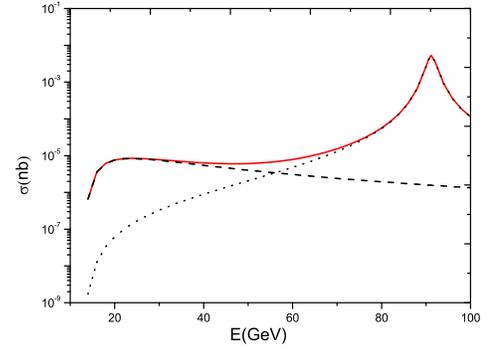}
\caption{Total cross section $\sigma(e^{+}+e^{-}\rightarrow\gamma^*/Z^0\rightarrow (c\bar{b})[n]+b+\bar{c})$ versus the $e^{+}e^{-}$ collision energy $E$, where contributions from the mentioned Fock states $|(c\bar{b})_{\bf 1,8}[^1S_0]\rangle$, $|(c\bar{b})_{\bf 1,8}[^3S_1]\rangle$, $|(c\bar{b})_{\bf 1}[^1P_1]\rangle$ and $|(c\bar{b})_{\bf 1}[^3P_J]\rangle$ have been summed up. Here the dashed, the dotted and the solid lines are for the production channel with $\gamma^*$-propagator, $Z^0$-propagator and the sum of these two channels, respectively.} \label{energy}
\end{figure}

Firstly, we calculate the total cross-section for the $e^+ e^-$ collider energy at the $Z^0$ peak, i.e.  $E=m_Z$. We obtain
\begin{eqnarray}
\sigma_{|(c\bar{b})_{\bf 1}[^1S_0]\rangle}&=&1.800 \;{\rm pb},\\
\sigma_{|(c\bar{b})_{\bf 1}[^3S_1]\rangle}&=&2.575 \;{\rm pb},\\
\sigma_{|(c\bar{b})_{\bf 1}[^1P_1]\rangle}&=&0.190 \;{\rm pb},\\
\sigma_{|(c\bar{b})_{\bf 1}[^3P_0]\rangle}&=&0.115 \;{\rm pb},\\
\sigma_{|(c\bar{b})_{\bf 1}[^3P_1]\rangle}&=&0.232 \;{\rm pb},\\
\sigma_{|(c\bar{b})_{\bf 1}[^3P_2]\rangle}&=&0.256 \;{\rm pb},\\
\sigma_{|(c\bar{b})_{\bf 8}[^1S_0]g\rangle}&=&0.225 \times{v^4}  \;{\rm pb},\\
\sigma_{|(c\bar{b})_{\bf 8}[^3S_1]g\rangle}&=&0.322 \times{v^4}  \;{\rm pb}.
\end{eqnarray}
It is found that the total cross-sections of all the $P$-wave states can be added up to $\sim20\%$ of that of the ground state $B_c(|(c\bar{b})_{\bf 1}[^1S_0])$. So the $P$-wave $(c\bar{b})$-quarkonium states should be taken into consideration, especially for the future high luminosity $e^+e^-$ colliders. Two color-octet $S$-wave states shall only provide $\sim 1\%$ contribution to the ground state.

\begin{figure}
\includegraphics[width=0.35\textwidth]{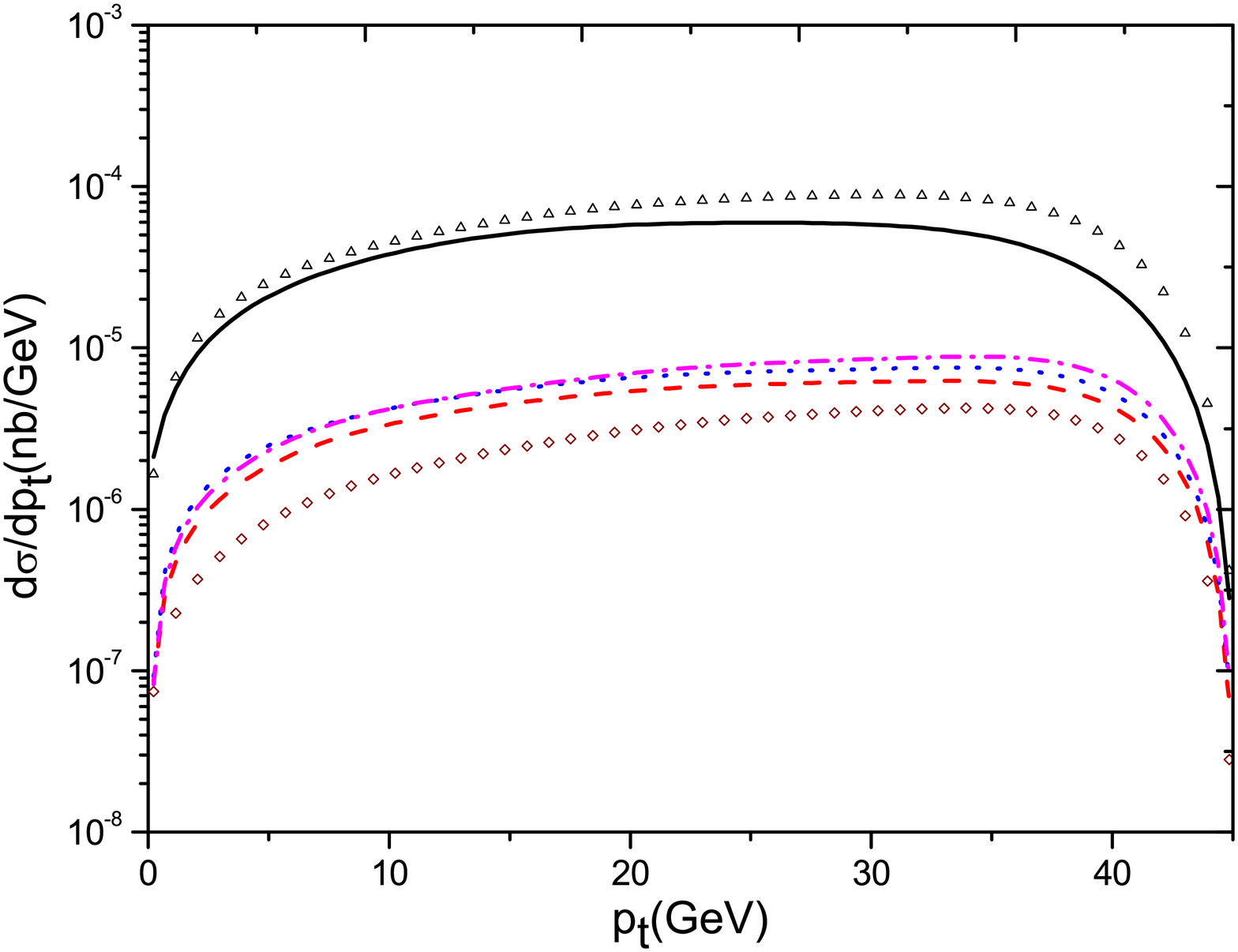}
\includegraphics[width=0.35\textwidth]{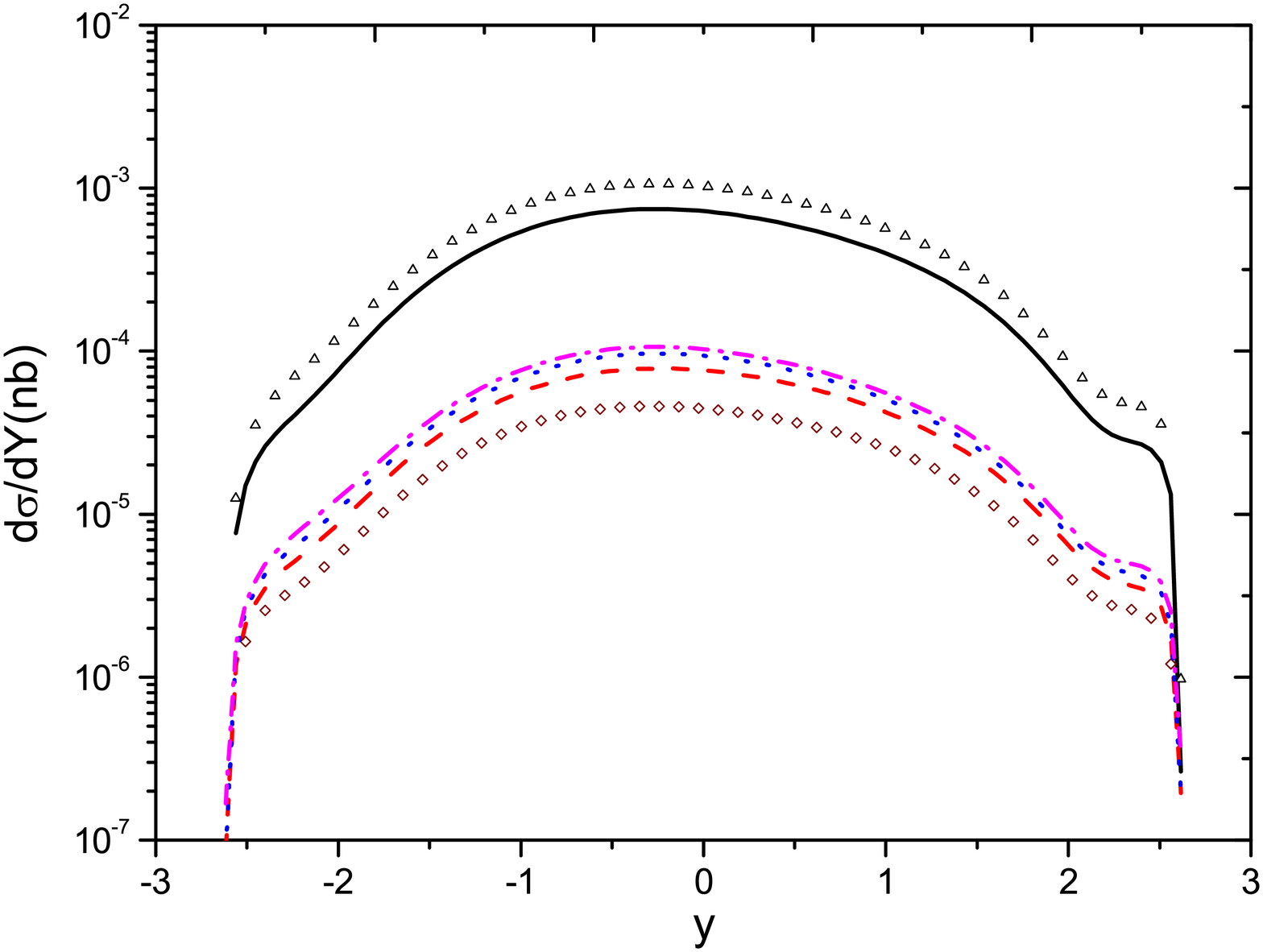}
\caption{The differential $p_t$- and $y$- distributions of the $(c\bar{b})$-quarkonium, where the solid, the dashed, the dotted, the dash-dot, the triangle and the diamond lines are for $|(c\bar{b})_{\bf 1}[^1S_0]\rangle$, $|(c\bar{b})_{\bf 1}[^1P_1]\rangle$, $|(c\bar{b})_{\bf 1}[^3P_1]\rangle$, $|(c\bar{b})_{\bf 1}[^3P_2]\rangle$, $|(c\bar{b})_{\bf 1}[^3S_1]\rangle$ and $|(c\bar{b})_{\bf 1}[^3P_0]\rangle$ respectively.} \label{ptrap}
\end{figure}

\begin{table}
\begin{tabular}{|c||c|c|c|c|c|}
\hline
  ~~~$E_{cm}$~~~    & ~$95\%\; m_Z$~& ~$97\%\; m_Z$~   & ~$m_Z$~   & ~$103\%\; m_Z$~ & ~$105\%\; m_Z$~ \\
\hline
$\sigma_{\rm tot}$ (pb) &  0.341  &  0.860   & 5.190  & 0.930  & 0.384 \\
\hline
\end{tabular}
\caption{Total cross-section of the process $e^{+}+e^{-} \rightarrow\gamma^*/Z^0\rightarrow (c\bar{b})[n]+b+\bar{c}$ under different collision energies, where contributions from the mentioned Fock states $|(c\bar{b})_{\bf 1,8}[^1S_0]\rangle$, $|(c\bar{b})_{\bf 1,8}[^3S_1]\rangle$, $|(c\bar{b})_{\bf 1}[^1P_1]\rangle$ and $|(c\bar{b})_{\bf 1}[^3P_J]\rangle$ have been summed up. Here $m_c=1.50$ GeV, $m_b=4.90$ GeV and $v^2=0.2$. } \label{diffmZ}
\end{table}

As a reference, we present the total cross section $\sigma(e^{+}+e^{-}\rightarrow\gamma^*/Z^0\rightarrow (c\bar{b})[n]+b+\bar{c})$ versus the $e^{+}e^{-}$ collision energy $E$ in Fig.(\ref{energy}), where contributions from the mentioned Fock states $|(c\bar{b})_{\bf 1,8}[^1S_0]\rangle$, $|(c\bar{b})_{\bf 1,8}[^3S_1]\rangle$, $|(c\bar{b})_{\bf 1}[^1P_1]\rangle$ and $|(c\bar{b})_{\bf 1}[^3P_J]\rangle$ have been summed up. We take $v^2=0.2$ for the color-octet states. For small $e^{+}e^{-}$ collision energy ($E\lesssim 60$ GeV), we have
\begin{displaymath}
\sigma_{e^{+}+e^{-}\to\gamma^*\to (c\bar{b})[n]} > \sigma_{e^{+}+e^{-}\to Z^0\to (c\bar{b})[n]} .
\end{displaymath}
One may observe a small peak around $20$ GeV, which is caused by the production channel through $\gamma^*$-propagator and has already been observed by Ref.\cite{zbc0}. When the collider energy runs from $\sim 10$ GeV to $100$ GeV, it is found that the total cross-section at the energy around $Z^0$-peak raises up to four orders than that of $10$ GeV due to the $Z^0$-resonance effects. Moreover, as shown by Table \ref{diffmZ}, if taking the collision energy $E=(1 \pm 3\%) m_Z$, the total cross-section shall be lowered to $\sim\left({}^{18\%}_{17\%}\right)$ of its peak value; and if taking the collision energy $E=(1 \pm 5\%) m_Z$, the total cross-section shall be lowered to $\sim 7\%$ of its peak value. Moreover, we present the $(c\bar{b})$-quarkonium transverse momentum ($p_t$)- and rapidity ($y$)- differential distributions in Fig.(\ref{ptrap}).

\begin{table}
\begin{tabular}{|c||c|c|c|c|c|}
\hline
~~$m_c$({\rm GeV})~~        & ~~1.20~~   & ~~1.35~~   & ~~1.50~~   & ~~1.65~~   & ~~1.80~~  \\
\hline \hline
$\sigma_{|(c\bar{b})_{\bf 1}[^1S_0]\rangle}({\rm pb})$ & 3.564  & 2.485  &  1.800  &  1.477  &  1.031 \\
\hline
$\sigma_{|(c\bar{b})_{\bf 1}[^3S_1]\rangle}({\rm pb})$ & 5.447  & 3.668  &  2.575  &  2.074  &  1.397 \\
\hline
$\sigma_{|(c\bar{b})_{\bf 1}[^1P_1]\rangle}({\rm pb})$ & 0.571  & 0.319  &  0.190  &  0.139  &  0.079 \\
\hline
$\sigma_{|(c\bar{b})_{\bf 1}[^3P_0]\rangle}({\rm pb})$ & 0.281  & 0.175  &  0.115  &  0.089  &  0.056  \\
\hline
$\sigma_{|(c\bar{b})_{\bf 1}[^3P_1]\rangle}({\rm pb})$ & 0.658  & 0.379  &  0.232  &  0.172  &  0.101  \\
\hline
$\sigma_{|(c\bar{b})_{\bf 1}[^3P_2]\rangle}({\rm pb})$ & 0.845  & 0.450  &  0.256  &  0.154  &  0.097  \\
\hline
\end{tabular}
\caption{Uncertainties for the total cross-section of the process $e^{+}+e^{-}\rightarrow (c\bar{b})[n]+b+\bar{c}$ with varying $m_c$, where $m_b$ is fixed to be $4.90$ GeV. }\label{section1}
\label{tabmc}
\end{table}

\begin{table}
\begin{tabular}{|c||c|c|c|c|c|}
\hline ~~$m_b$ ({\rm GeV})~~    & ~~4.50~~   & ~~4.70~~   & ~~4.90~~   & ~~5.10~~   & ~~5.30~~  \\
\hline \hline
$\sigma_{|(c\bar{b})_{\bf 1}[^1S_0]\rangle}({\rm pb})$ & 1.816  & 1.808  &  1.800  & 1.793  &  1.786 \\
\hline
$\sigma_{|(c\bar{b})_{\bf 1}[^3S_1]\rangle}({\rm pb})$ & 2.526  & 2.551  &  2.575  & 2.599  &  2.623 \\
\hline
$\sigma_{|(c\bar{b})_{\bf 1}[^1P_1]\rangle}({\rm pb})$ & 0.195  & 0.192  &  0.190  & 0.188  &  0.186  \\
\hline
$\sigma_{|(c\bar{b})_{\bf 1}[^3P_0]\rangle}({\rm pb})$ & 0.126  & 0.120  &  0.115  & 0.110  &  0.105  \\
\hline
$\sigma_{|(c\bar{b})_{\bf 1}[^3P_1]\rangle}({\rm pb})$ & 0.243  & 0.237  &  0.232  & 0.227  &  0.223  \\
\hline
$\sigma_{|(c\bar{b})_{\bf 1}[^3P_2]\rangle}({\rm pb})$ & 0.252  & 0.254  &  0.256  & 0.258  &  0.260  \\
\hline
\end{tabular}
\caption{Uncertainties for the total cross-section of the process $e^{+}+e^{-}\rightarrow (c\bar{b})[n]+b+\bar{c}$ with varying $m_b$, where $m_c$ is fixed to be $1.50$ GeV. }\label{section2}
\label{tabmb}
\end{table}

\begin{figure}
\includegraphics[width=0.35\textwidth]{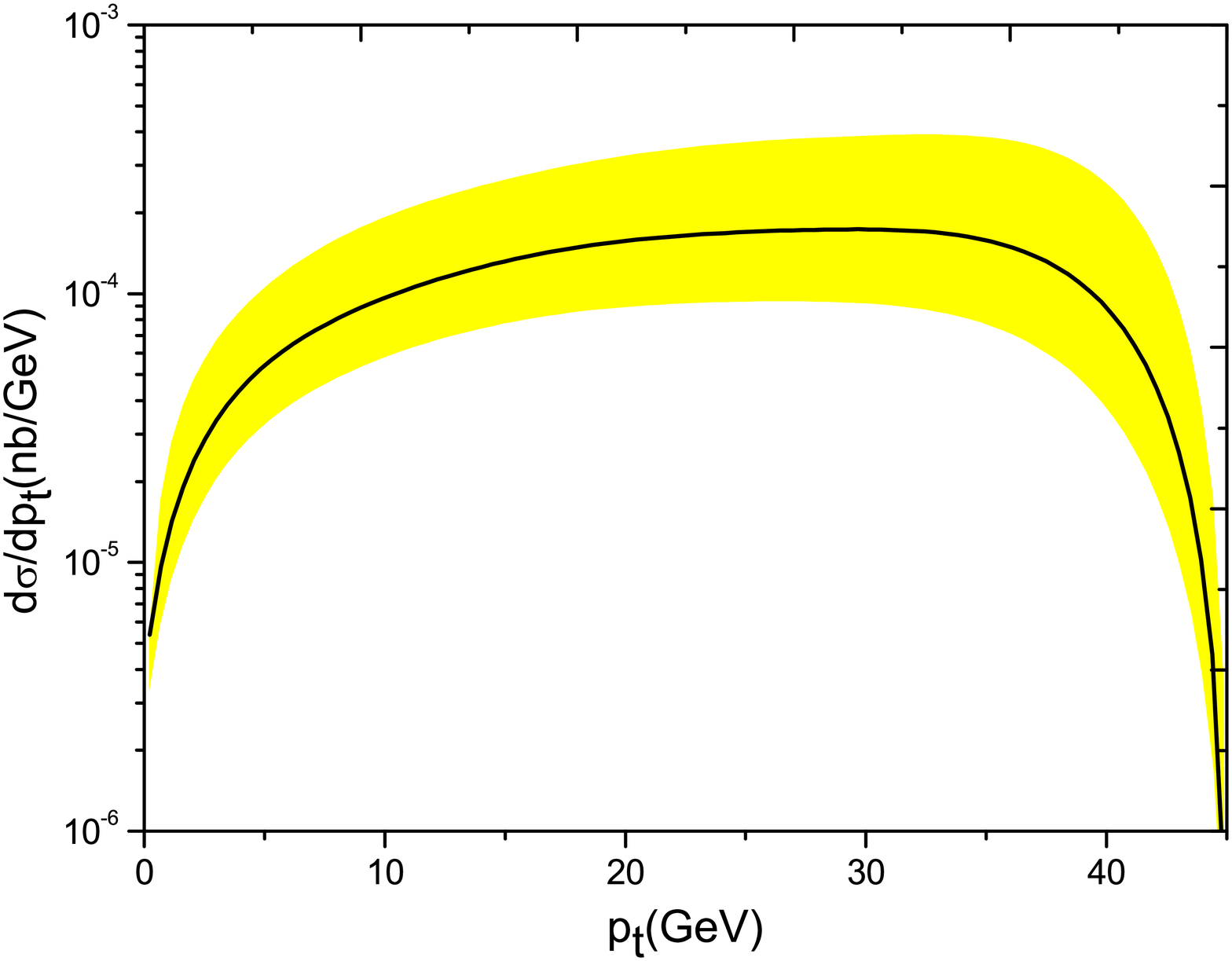}
\includegraphics[width=0.35\textwidth]{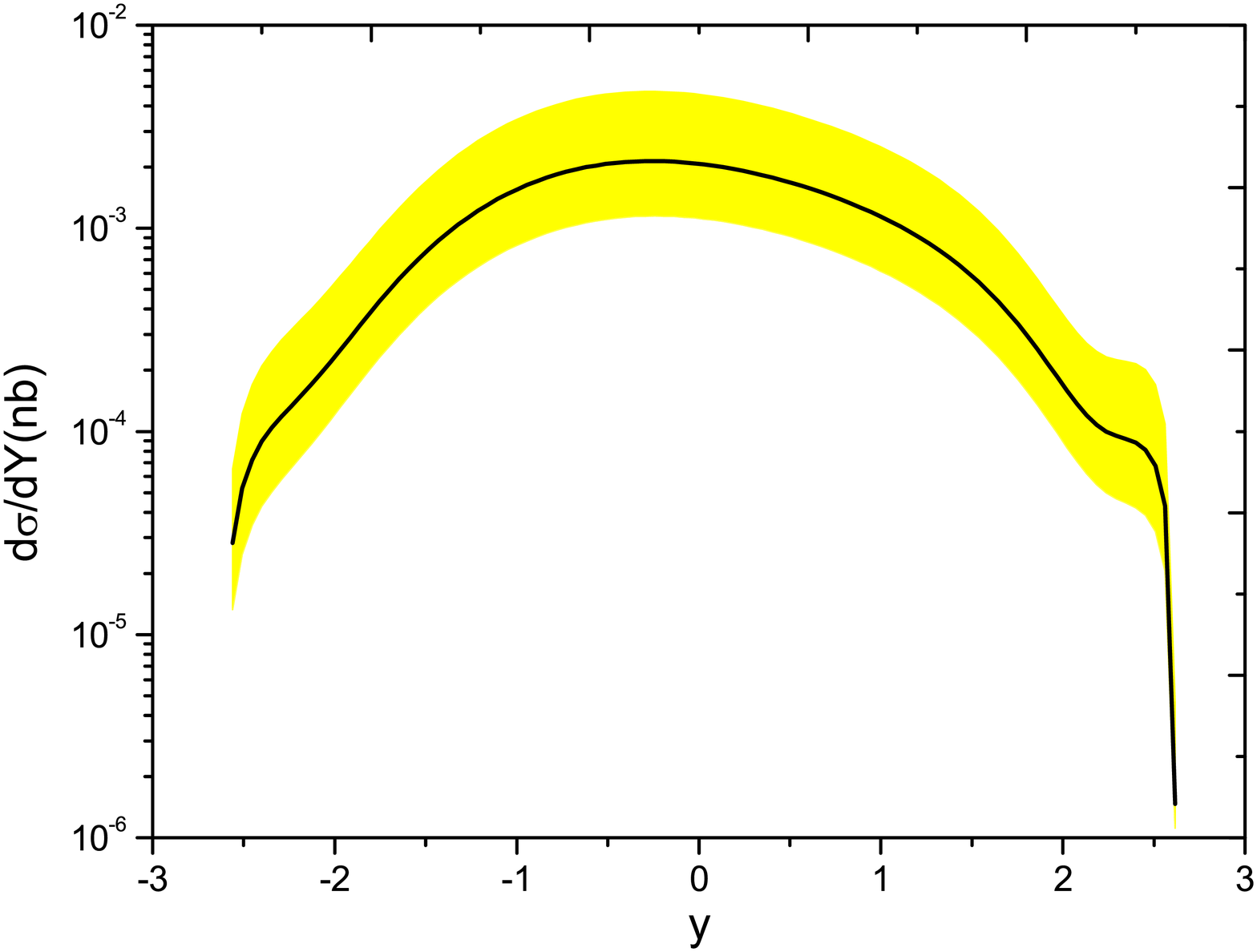}
\caption{Uncertainties of $p_t$- and $y$- differential distributions for the $(c\bar{b})$-quarkonium, where contributions from the Fock states $|(c\bar{b})_{\bf 1,8}[^1S_0]\rangle$, $|(c\bar{b})_{\bf 1,8}[^3S_1]\rangle$, $|(c\bar{b})_{\bf 1}[^1P_1]\rangle$ and $|(c\bar{b})_{\bf 1}[^3P_J]\rangle$ have been summed up.} \label{ptrapun}
\end{figure}

Secondly, we discuss the theoretical uncertainties caused by the $c$ and $b$ quarks by varying $m_c=1.50\pm0.30$ GeV and $m_b=4.90\pm0.40$ GeV. There are other uncertainty sources include the matrix elements, the renormalization scale $\mu_R$ and etc. There are overall factors, their uncertainties can be easily estimated when their values are known, so we shall not discuss them there. The cross-section for the $(c\bar{b})$-quarkonium production through $e^+e^-$ annihilation with varying $m_c$ and $m_b$ are presented in Tabs.({\ref{section1}},{\ref{section2}}). It can be found that the total cross-section is more sensitive to $m_c$ than that of $m_b$. And the total cross-section of the $P$-wave states are more sensitive to quark masses than the $S$-wave states. By adding the uncertainties in quadrature, we obtain
\begin{eqnarray}
\sigma_{|(c\bar{b})_{\bf 1}[^1S_0]\rangle}&=&1.800^{+1.764}_{-0.769} \;{\rm pb},\\
\sigma_{|(c\bar{b})_{\bf 1}[^3S_1]\rangle}&=&2.575^{+2.872}_{-1.179} \;{\rm pb},\\
\sigma_{|(c\bar{b})_{\bf 1}[^1P_1]\rangle}&=&0.190^{+0.381}_{-0.111} \;{\rm pb},\\
\sigma_{|(c\bar{b})_{\bf 1}[^3P_0]\rangle}&=&0.115^{+0.166}_{-0.060} \;{\rm pb},\\
\sigma_{|(c\bar{b})_{\bf 1}[^3P_1]\rangle}&=&0.232^{+0.426}_{-0.131} \;{\rm pb},\\
\sigma_{|(c\bar{b})_{\bf 1}[^3P_2]\rangle}&=&0.256^{+0.589}_{-0.159} \;{\rm pb},\\
\sigma_{|(c\bar{b})_{\bf 8}[^1S_0]g\rangle}&=&0.225^{+0.221}_{-0.096} \times{v^4}  \;{\rm pb},\\
\sigma_{|(c\bar{b})_{\bf 8}[^3S_1]g\rangle}&=&0.322^{+0.359}_{-0.147} \times{v^4}  \;{\rm pb}.
\end{eqnarray}
If assuming the higher excited states decay to the ground state $B_c(|(c\bar{b}_{\bf 1})[^1S_0]\rangle)$ with 100\% efficiency, then we obtain the total cross-section for $B_c$, i.e.
\begin{equation}
\sigma_{e^{+}+e^{-}\rightarrow B_c +b+\bar{c}}=5.190^{+6.222}_{-2.419} \;{\rm pb}.
\end{equation}
Furthermore, to show the corresponding uncertainties more clearly, we present the uncertainties of $p_t$- and $y$- differential distributions for the $(c\bar{b})$-quarkonium production in Fig.(\ref{ptrapun}), where the Fock states $|(c\bar{b})_{\bf 1,8}[^1S_0]\rangle$, $|(c\bar{b})_{\bf 1,8}[^3S_1]\rangle$, $|(c\bar{b})_{\bf 1}[^1P_1]\rangle$ and $|(c\bar{b})_{\bf 1}[^3P_J]\rangle$ states have been summed up. The uncertainties are drawn by a shaded band, where the center solid line is for $m_c=1.5$ GeV and $m_b=4.9$ GeV, the upper edge of the band is for $m_c=1.2$ GeV and $m_b=5.3$ GeV, and the lower edge of the band is for $m_c=1.8$ GeV and $m_b=4.5$ GeV.

\begin{figure}
\includegraphics[width=0.35\textwidth]{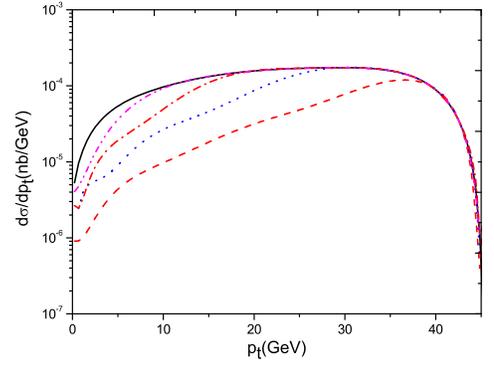}
\caption{The $(c\bar{b})$-quarkonium differential distributions versus $p_t$ with various $y_{cut}$, where contributions from all the considered Fock states have been summed up. The solid, the dashed, the dotted, the dash-dot and the dash-dot-dot lines are for $y_{cut}=0$, $0.5$, $1.0$, $1.5$ and $2.0$ respectively. } \label{ptvaryrap}
\end{figure}

\begin{figure}
\includegraphics[width=0.35\textwidth]{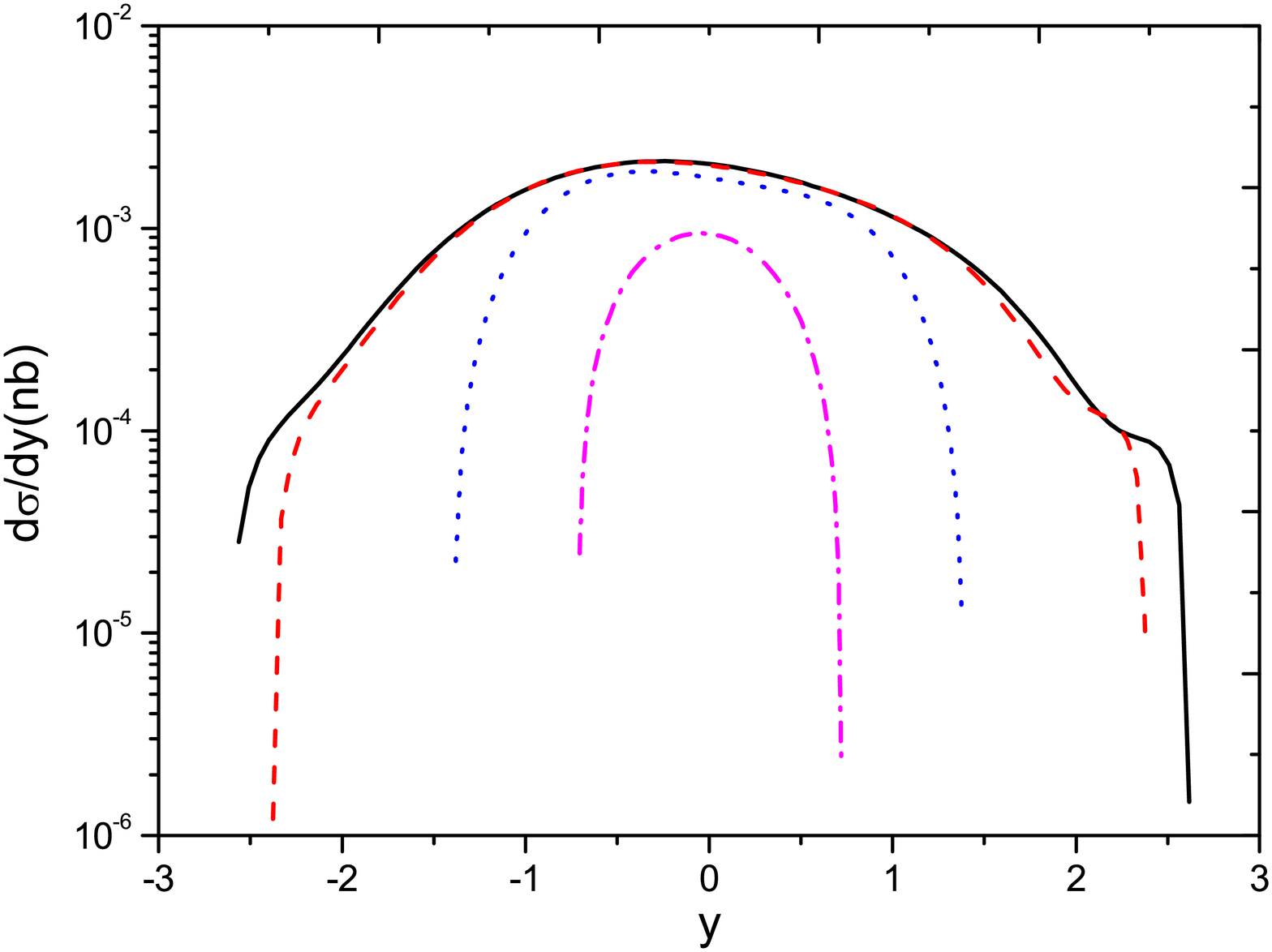}
\caption{The $(c\bar{b})$-quarkonium differential distributions versus $y$ with various $p_{tcut}$, where contributions from all the considered Fock states have been summed up. The solid, the dashed, the doted and the dash-dot lines are for $p_{tcut}=0$, $5$, $20$, $35$ GeV respectively. } \label{rapvarypt}
\end{figure}

\begin{table}[ht]
\begin{center}
\caption{$R_{p_{tcut}}$ versus the parameters $p_{tcut}$ and the $y_{cut}$.} \vskip 0.6cm
\begin{tabular}{|c||c|c|c||c|c|c||c|c|c|}
\hline\hline $p_{tcut}$ & \multicolumn{3}{|c||}{$0$ GeV}&
\multicolumn{3}{|c||}{$5$ GeV}& \multicolumn{3}{|c|}{$20$ GeV}\\
\hline $y_{cut}$& 0.5 & 1.0 & 1.5 & 0.5 & 1.0 & 1.5 & 0.5 & 1.0 & 1.5\\
\hline\hline $R_{p_{tcut}}$ & 0.382 & 0.710 & 0.897 & 0.391 & 0.725 & 0.913 & 0.520 & 0.913 & 1.00 \\
\hline\hline
\end{tabular}
\label{tevcut}
\end{center}
\end{table}

Experimentally, there is no detector which can cover all the kinematics of the events, so only some of the $B_c$ events can be observed. That is, the $B_c$ events with small $p_t$ and/or a large rapidity $y$ cannot be detected, so they cannot be utilized for experimental studies. As a final remark, it would be helpful to know how the differential cross-sections change with different $p_t$- or $y$- cuts. In Fig.(\ref{ptvaryrap}), we present the $(c\bar{b})$-quarkonium $p_t$-distributions for all the considered Fock states with a particular rapidity cut ($y_{cut}$), where the solid, the dashed, the dotted, the dash-dot and the dash-dot-dot lines are for $y_{cut}=0$, $0.5$, $1.0$, $1.5$ and $2.0$ respectively. In Fig.(\ref{rapvarypt}), we present the $(c\bar{b})$-quarkonium $y$-distributions for all the considered Fock states with a particular transverse momentum cut ($p_{tcut}$), where the solid, the dashed, the doted and the dash-dot lines are for $p_{tcut}=0, 5, 20, 35$ GeV respectively. To analyze the quantitative difference of the differential distributions with regard to the parameters $p_{tcut}$ and $y_{cut}$, we introduce a ratio for the integrated total cross-section:
\begin{equation}
R_{p_{tcut}}=\left(\frac{\sigma_{y_{cut}}}{\sigma_{0}} \right)_{p_{Tcut}} ,
\end{equation}
where $\sigma_{y_{cut}}$ and $\sigma_{0}$ are the total cross-section with and without $y_{cut}$ respectively. In the definition of $R_{p_{tcut}}$, contributions from all the considered Fock states have been summed up. The values of the ratio $R_{p_{tcut}}$ that correspond to different values of $p_{tcut}$ and $y_{cut}$ are presented in Tab.(\ref{tevcut}). It is found that when under the same $p_{tcut}$, $R_{p_{tcut}}$ increases with the increment of $y_{cut}$, and when under the same $y_{cut}$, $R_{p_{tcut}}$ also increases with $p_{tcut}$. This shows that the rapidity distribution of the  $(c\bar{b})$-quarkonium shrinks with the increment of $p_{tcut}$; e.g. when $p_{tcut}=20 $ GeV, $|y|\lesssim 1.5$.

\section{summary}

In the present paper, we have studied the $(c\bar{b})$-quarkonium production through $e^+e^-$ annihilation in leading $\alpha_s$ order, but with the $v^2$-expansion up to $v^4$. The production of the $(c\bar{b})$-quarkonium in Fock states $|(c\bar{b})_{\bf 1}[^1S_0]\rangle$, $|(c\bar{b})_{\bf 1}[^3S_1]\rangle$, $|(c\bar{b})_{\bf 1}[^1P_1]\rangle$, $|(c\bar{b})_{\bf 1}[^3P_J]\rangle$, $|(c\bar{b})_{\bf 8}[^1S_0]g\rangle$ and $|(c\bar{b})_{\bf 8}[^3S_1]g\rangle$ have been studied. Higher Fock states such as the $P$-wave states will also provide sizable contributions to the total cross-section of the $B_c$ meson. If assuming the higher excited states decay to the ground state $B_c(|(c\bar{b}_{\bf 1})[^1S_0]\rangle)$ with 100\% efficiency via electromagnetic or hadronic interactions, then we will obtain $\sigma_{e^{+}+e^{-}\rightarrow B_c +b+\bar{c}}=5.190^{+6.222}_{-2.419}$ pb.

According to our present calculation, one would expect to accumulate $10^5 \sim 10^7$ $B_c$ events in one operational year for the super Z-factory with a high luminosity ${\cal L}\propto 10^{34} \sim10^{36} cm^{-2}s^{-1}$. If taking the $e^+ e^-$ collision energy to run slightly off the $Z^0$-peak; i.e. $\sqrt{S}=0.95 m_Z$ or $1.05 m_Z$, the total production cross-section will be lowered by about one-order from its peak value. However one can estimate $10^4 - 10^6$ $B_c$ events to be produced under such condition, which is still sizable. Then, in addition to the hadronic colliders Tevatron and LHC, the super Z-factory will provide another good platform for studying the $B_c$-meson properties.

\vspace{20mm} \noindent
{\large\bf Acknowledgements:}
This work was supported in part by the Fundamental Research Funds for the Central Universities under Grant No. CDJXS1102209, the Program for New Century Excellent Talents in University under Grant No. NCET-10-0882 and by Natural Science Foundation of China under Grant No.10805082 and No.11075225.\\

\appendix

\section{Linear polarization vector and tensor}

We set the $z$-axis to be along the incoming beam. The momentum of the $(c\bar{b})$-quarkonium $q_3^\mu=(q^0_3,|\vec{q_3}|\sin\theta\cos\phi,|\vec{q_3}|\sin\theta\sin\phi,|\vec{q_3}|\cos\theta)$, where $|\vec{q_3}|=\sqrt{(q_3^0)^2-M^2}$, $M$ is the quarkonium mass, $\theta$ and $\phi$ are polar and azimuth angles respectively. The expressions of the polarization vector can be written as
\begin{eqnarray}
\epsilon_{x}(\vec{q}_3)&=&(0, \cos\theta\cos\phi, \cos\theta\sin\phi, -\sin\theta), \nonumber\\
\epsilon_{y}(\vec{q}_3)&=&(0, -\sin\phi, \cos\phi, 0), \nonumber\\
\epsilon_{z}(\vec{q}_3)&=&\frac{1}{M}(|\vec{q_3}|, q_3^0\sin\theta\cos\phi,
q_3^0\sin\theta\sin\phi, q_3^0\cos\theta) \nonumber .
\end{eqnarray}
As required, the polarization vector satisfies the following relation
\begin{equation}
\epsilon_{i}\cdot q_3=0\,,\,\;\;\; \epsilon_{i} \cdot
\epsilon_{j}=-\delta_{ij}\,,\,\,\;\;\;(i,j=x,y,z)\,.
\end{equation}

The polarization tensor $\epsilon^{\alpha\beta}_{J_z}(q_3)$ for $(^{3}P_{2})$-quarkonium state can be constructed in terms of $\epsilon_{x}(\vec{q_3})$, $\epsilon_{y}(\vec{q_3})$ and $\epsilon_{z}(\vec{q_3})$ as
\begin{eqnarray}
&\epsilon^{\mu\nu}_{1}=\frac{1}{\sqrt{2}}(\epsilon_x^{\mu}
\epsilon_y^{\nu}+\epsilon_y^{\mu} \epsilon_x^{\nu})\,,\;\;\;\;
\epsilon^{\mu\nu}_{2}=\frac{1}{\sqrt{2}}(\epsilon_x^{\mu}
\epsilon_z^{\nu}+\epsilon_z^{\mu} \epsilon_x^{\nu})\,,\nonumber \\
&\epsilon^{\mu\nu}_{3}=\frac{1}{\sqrt{2}}(\epsilon_y^{\mu}
\epsilon_z^{\nu}+\epsilon_z^{\mu} \epsilon_y^{\nu})\,,\;\;\;\;
\epsilon^{\mu\nu}_{4}=\frac{1}{\sqrt{2}}(\epsilon_x^{\mu}
\epsilon_x^{\nu}-\epsilon_y^{\mu} \epsilon_y^{\nu})\,,\nonumber\\
&\epsilon^{\mu\nu}_{5}=\frac{1}{\sqrt{6}}(\epsilon_x^{\mu}
\epsilon_x^{\nu}+\epsilon_y^{\mu} \epsilon_y^{\nu}-
2\epsilon_z^{\mu} \epsilon_z^{\nu})\;.
\end{eqnarray}

\section{New lorentz structures and their coefficients for $e^{+} +e^{-}\rightarrow Z^0 \rightarrow (c\bar{b})[n] +b+\bar{c}$}

In the present section, we present the new lorentz structures and their non-zero coefficients for the process $e^{+}(p_2) +e^{-}(p_1)\rightarrow Z^0/\gamma^* \rightarrow (c\bar{b})[n] (q_3)+b(q_2)+\bar{c}(q_1)$, where $n$ stands for $|(c\bar{b})_{\bf 1}[^1S_0]\rangle$, $|(c\bar{b})_{\bf 1}[^3S_1]\rangle$, $|(c\bar{b})_{\bf 1}[^1P_1]\rangle$, $|(c\bar{b})_{\bf 1}[^3P_J]\rangle$(J=1,2,3) respectively. All these new Lorentz structures have no correspondence to the Lorentz structures introduced from the process $Z^{0}(k)\rightarrow (c\bar{b})[n](q_3) +b(q_2)+\bar{c}(q_1)$. Note, the following parameters, $r_1$, $r_2$, $r_3$, $T_b$, $T_c$ and so on, have the same meaning as those of Refs.\cite{zbc1,zbc2}, and we shall not present them here for short.

\subsection{Spin-singlet S-wave state: $|(c\bar{b})[^{1}S_0]\rangle$}

There is only one new Lorentz structure for $|(c\bar{b})[^{1}S_0]\rangle$,
\begin{eqnarray}
B_1=\frac{k^{\nu}}{m_Z} ,
\end{eqnarray}
whose coefficients $A^{1,2,3',4'}_1$ are
\begin{widetext}
\begin{eqnarray}
A^1_1 &=& \frac{{m_Z}^{7/2}{L_1}}{\sqrt{r_3}} (d_4 r_1 (3 r_1-r_2) r_3 +d_1 (3 r_3 {r_1}^2
+(-r_2 r_3+u-2 x) r_1-r_2 u) \nonumber\\
&&+(d_2 (r_1 (3 r_1+r_2)+u-2 y)+d_3 (r_1 (3 r_1+r_2)+3 u-2 y)) r_3+d_4 (3 r_1+r_2) u),\\
A^2_1 &=&\frac{4 {m_Z}^{7/2}{L_2}}{\sqrt{r_3}} (((d_3+d_4) T_c-d_2 T_b) (r_1 (3 r_1-r_2) r_3+(3 r_1+r_2) u)+ d_1 T_b (3 r_3 r_1^2-(r_2 r_3  \nonumber\\
&&-3 u+2 x) r_1+r_2 u)+2 r_3 (d_2 T_b-d_3 T_c) y)  \\
A^{3'}_1 &=&\frac{{m_Z}^{9/2} N_0}{\sqrt{r_3} L_2} (d_1 T_b (u (2 y-1)+2 r_1 r_3 (x+2 y-1))
+d_2 T_b (u-2 (-r_2 r_3 y+r_1 r_3 (x+y-1) \nonumber\\
&&+u (x+y)))+T_c (2 d_4 (r_1 r_3+u) (2 r_1 r_3+u)+d_3 (u (2 y-1)+2 r_1 r_3 (x+2 y-1))))\nonumber\\
A^{4'}_1 &=&-\frac{1}{4 \sqrt{r_3} L_1}({m_Z}^{9/2} N_0 (2 d_4 (2 r_1^4+r_2 r_1^3+3 u r_1^2
+r_2^3 r_1+u (u-r_2^2))-d_3 (-4 y r_3^2  \nonumber\\
&&+2 (r_2 r_1^2-(r_2^2+x-1) r_1+r_2 u) r_3+u-2 u y)+d_2 (u+2 (r_1^4+(-r_2^2+u+x \nonumber\\
&&+y-1) r_1^2+r_2 (u+x+2 y-1) r_1+r_2^2 y-u (x+y)))+d_1 (6 r_3 r_1^3 \nonumber\\
&&-2 (-3 u+3 x+2 y-1) r_1^2-2 r_3 r_2 r_1^2+2 r_2 (u-x-2 y+1) r_1+u (2 y-1))))
\end{eqnarray}
\end{widetext}

\subsection{Spin-triplet S-wave state: $|(c\bar{b})[^3S_1]\rangle$}

There are three new Lorentz structures for $|(c\bar{b})[^3S_1]\rangle$,
\begin{eqnarray}
B_1 &=&\frac{1}{m_Z^2}k\cdot\epsilon(q_3)k^{\nu},\;\;
B_2 =\frac{1}{m_Z^2}q_2\cdot\epsilon(q_3)k^{\nu},\;\;\nonumber\\
B_3 &=& \frac{i}{m_Z^4} \varepsilon(k,q_3,q_2,\epsilon(q_3))k^{\nu} ,
\end{eqnarray}
whose coefficients $A^{1,2,3',4'}_j$ ($j=1,2,3$) are
\begin{eqnarray}
A^1_1 &=& 8 d_1 L_1 {m_Z}^{7/2} {r_1} \sqrt{r_3} T_b,\\
A^1_2 &=& 4 L_1 {m_Z}^{7/2} r_3^{3/2} [(d_2-d_1)T_b -(d_3+d_4)T_c],\\
A^1_3 &=& -\frac{2 L_1 {m_Z}^{7/2}}{\sqrt{r_3}}(d_2-d_3),\\
A^2_1 &=&2 d_1 L_2 {m_Z}^{7/2} r_1 \sqrt{r_3},\\
A^2_2 &=&-L_2 {m_Z}^{7/2}\sqrt{r_3} [(3 d_1-d_2+d_3+d_4) r_1 \nonumber\\
&&-(d_1-d_2+d_3-d_4) r_2] ,\\
A^2_3 &=&\frac{8 L_2 {m_Z}^{7/2}}{\sqrt{r_3}} (d_2 T_b+d_3 T_c),\\
A^{3'}_1 &=& \frac{d_1 {m_Z}^{9/2} N_0 \sqrt{r_3}}{2 L_2}(r_1 r_3+u),\\
A^{3'}_2 &=& -\frac{{m_Z}^{9/2} N_0 \sqrt{r_3}}{4 L_2}(d_2-d_3+2 d_4 ({r_3}^2+ u) \nonumber\\
&&+d_1 (-1+2 r1 r_3+2 x+2 y)+\nonumber\\
&&2 (d_2 (r_1 r_3-x-y)+d_3 (r_2 r_3+y))),\\
A^{3'}_3 &=&\frac{2 {m_Z}^{9/2} N_0 \sqrt{r_3}}{L_2}(d_1 T_b+d_3 T_c) ,\\
A^{4'}_1 &=&-\frac{2 d_1 {m_Z}^{9/2} N_0 \sqrt{r_3} T_b}{L_1} (r_1^2-r_2 r_1+u),\\
A^{4'}_2 &=&\frac{{m_Z}^{9/2} N_0 \sqrt{r_3}}{L_1}(d_2 T_b (-2 x-2 y+1)+\nonumber\\
&& d_1 T_b (2 x+2 y-1)+ T_c (d_4 (r_1^2-r_2^2+r_3^2 \nonumber\\
&&+2 u)+d_3 (2 y-1))),\\
A^{4'}_3 &=&\frac{{m_Z}^{9/2} N_0}{2 L_1 \sqrt{r_3}} [(d_2-d_3) (r_1-r_2)+(d_2-d_1) r_3]
\end{eqnarray}

\subsection{Spin-singlet P-wave state: $|(c\bar{b})[^1P_1]\rangle$}

There are three new Lorentz structure for $|(c\bar{b})[^1P_1]\rangle$, which are the same as that of $(c\bar{b})[^3S_1]$, whose coefficients $A^{1,2,3',4'}_j$ ($j=1,2,3$) are
\begin{widetext}
\begin{eqnarray}
A^1_1 &=&\frac{{m_Z}^{5/2} L_1}{r_2 \sqrt{r_3}}(2 d_{11} r_3 r_1 r_2^2-2 ((d_{21}+d_{22}
-d_{32}) r_1 (3 r_1+r_2) r_3+(d_{21}+d_{22}-3 d_{32}) u r_3 \nonumber\\
&&+d_{11} (3 r_3 r_1^2+u r_1-2 x r_1-r_2 u)) r_2+4 (d_{21}+d_{22}
-d_{32}) r_3 y r_2+d_1 (r_2-r_1) r_3),\\
A^1_2 &=&\frac{1}{2 r_1 r_2 \sqrt{r_3}}({m_Z}^{5/2} L_1 (-d_1 r_3 r_2^2+d_4 r_3 r_2^2
+d_2 r_3^2 r_2-d_3 r_3^2 r_2-8 (d_{22}+d_{31}) r_1 r_3 y r_2 \nonumber\\
&&+4 d_{11} r_1 (3 r_3 r_1^2+(u-2 x) r_1-r_2 u-r_2 r_3 r_1) r_2+r_1 (-4 d_{41} r_3 r_1 r_2^2 \nonumber\\
&&+4 (d_{41} (3 r_1+r_2)+(d_{22}+3 d_{31}) r_3) u r_2+r_3 (4 (d_{22}+d_{31}) r_1 r_2^2 \nonumber\\
&&-2 (-6 (d_{22}+d_{31}+d_{41}) r_1^2+d_1+d_4) r_2+(3 d_1+d_4) r_1+(d_3-d_2) r_3)))),\\
A^1_3 &=&-\frac{4 L_1 {m_Z}^{5/2} \sqrt{{r_3}}}{{r_1} {r_2}}({d_2} {T_b}+{d_3} {T_c}),\\
A^2_1 &=&\frac{4 {m_Z}^{5/2} L_2}{r_2 \sqrt{r_3}}(d_1 (r_2^2-r_1^2) T_b
+2 r_2 (((d_{21}+d_{22}) T_b+d_{32} T_c) (3 r_3 r_1^2+3 u r_1+r_2 u-2 r_3 y \nonumber\\
&&-r_2 r_3 r_1)-d_{11} T_b (3 r_3 r_1^2+3 u r_1-2 x r_1+r_2 u-r_2 r_3 r_1))),\\
A^2_2 &=&\frac{1}{r_1 r_2 \sqrt{r_3}}(2 L_2 {m_Z}^{5/2} (d_1 r_3 T_b (r_1-r_2)^2
-d_2 r_3 T_b (r_1-r_2)^2+d_3 r_3 T_c (r_1-r_2)^2+d_4 r_3 T_c (r_1  \nonumber\\
&&-r_2)^2+4 d_{11} r_1 r_2 T_b (3 r_3 r_1^2+3 u r_1-2 x r_1+r_2 u-r_2 r_3 r_1)
-4 d_{22} r_1 r_2 T_b (3 r_3 r_1^2+3 u r_1  \nonumber\\
&&+r_2 u-2 r_3 y-r_2 r_3 r_1)+4 d_{41} r_1 r_2 T_c (3 r_3 r_1^2+3 u r_1+r_2 u-r_2 r_3) \nonumber\\
&&+4 d_{31} r_1 r_2 T_c (3 r_3 r_1^2+3 u r_1+r_2 u-2 r_3 y-r_2 r_3))),\\
A^2_3 &=&\frac{ {m_Z}^{5/2} \sqrt{r_3} L_2}{r_1 r_2}(d_2-d_3),\\
A^{3'}_1 &=&\frac{1}{L_2 r_1 r_2 \sqrt{r_3}}({m_Z}^{7/2} N_0 (d_1 (r_2-r_1) r_3 T_b (r_1 r_3+u)
+2 d_{21} r_1 r_2 T_b (2 r_1 r_3 (x-1)+u (2 x-1))  \nonumber\\
&&+2 d_{22} r_1 r_2 T_b (2 r_1 r_3 (x-1)+u (2 x-1))+4 (d_{21}+d_{22}) r_1
(r_1-r_2) r_2 r_3 T_b y+4 (d_{21}+d_{22}) r_1 r_2 \nonumber\\
&&\times T_b u y+2 d_{11} r_1 r_2 T_b (-2 y u+u-2 r_1 r_3 (x+2 y-1)) \nonumber\\
&&+2 d_{32} r_1 r_2 T_c (u (2 y-1)+2 r_1 r_3 (x+2 y-1)))),\\
A^{3'}_2 &=&\frac{1}{2 L_2 r_1 r_2 \sqrt{r_3}}({m_Z}^{7/2} N_0 (d_3 r_3 (r_3 (r_3-r_2) r_2
+r_2+r_1 (r_2 r_3-1)) T_c+d_4 (r_1-r_2) r_3 (r_1^2-r_2^2 \nonumber\\
&&+r_3^2+2 u) T_c+4 d_{41} r_1 r_2 (2 r_1 r_3+u) (r_1^2-r_2^2
+r_3^2+2 u) T_c+4 d_{31} r_1 r_2 (2 r_1 r_3 (x-1)-u) T_c  \nonumber\\
&&+2 d_{32} r_1 r_2 r_3 (2 r_2 x-r_3) T_c+2 (d_3 (r_1-r_2) r_3
+4 d_{31} r_1 r_2 (2 r_1 r_3+u)) y T_c+2 d_{21} r_1 r_2 r_3 T_b (r_3  \nonumber\\
&&-2 r_1 x)+d_1 (r_1-r_2) r_3 T_b (2 x+2 y-1)-4 d_{11} r_1 r_2 T_b (-2 y u+u-2 r_1 r_3 (x+2 y-1))  \nonumber\\
&&-d_2 r_3 T_b (2 r_2 r_1^2+(2 r_2^2+2 x+2 y-1) r_1+r_2-2 r_2 (x+y))  \nonumber\\
&&+4 d_{22} r_1 r_2 T_b (u-2 (-r_2 r_3 y+r_1 r_3 (x+y-1)+u (x+y))))),\\
A^{3'}_3 &=&-\frac{{m_Z}^{7/2} N_0  \sqrt{r_3}}{4 r_1 r_2 L_2}(-2 d_2 r_1+d_1 (r_1-r_2)+d_3 r_3) \\
A^{4'}_1 &=&\frac{{m_Z}^{7/2} N_0}{4 r_1 r_2 \sqrt{r_3} L_1}(-4 d_{11} r_2 r_3 r_2 r_1^3
+2 d_{11} r_2 (-u+2 r_1 (r_3+(3 r_1+r_2) (u-x)+3) \nonumber\\
&&+2 (u-2 r_1 r_3) y) r_1+2 r_2 (d_{32} (u+2 r_3 (r_2 r_1^2-(r_2^2+x-1) r_1
+r_2 u)-2 (2 r_3^2+u) y)  \nonumber\\
&&+d_{21} (u+2 (r_3 r_1^3-r_2 r_3 r_1^2+r_3 (u+x-1) r_1+r_3^2 y-u (x+y)))
+d_{22} (u+2 (r_3 r_1^3 \nonumber\\
&&-r_2 r_3 r_1^2+r_3 (u+x-1) r_1+r_3^2 y-u (x+y)))) r_1+d_1 (r_1-r_2) r_3 (r_1^2-r_2 r_1+u)), \\
A^{4'}_2 &=&\frac{{m_Z}^{7/2} N_0}{8 L_1 r_1 r_2 \sqrt{r_3}}(-2 d_4 (r_1-r_2) r_3 (r_1^2
+r_2^2+u)-8 d_{41} r_1 r_2 (2 r_1^4+r_2 r_1^3+3 u r_1^2+r_2^3 r_1+u (u  \nonumber\\
&&-r_2^2))+2 d_{21} r_1 r_2 r_3 (2 r_1 x-r_3)+2 d_{32} r_1 r_2 r_3 (r_3-2 r_2 x)
-d_1 (r_1-r_2) r_3(2 r_1(r_1-r_2)+2 x  \nonumber\\
&&+2 y-1)+4 d_{31} r_1 r_2 (u+2 r_3 (r_2 r_1^2
-(r_2^2+x-1) r_1+r_2 u)-2 (2 r_3^2+u) y)-d_2 r_3 (2 r_3 r_1^2 \nonumber\\
&&+(-2 x-2 y+1) r_1-r_2+2 r_2 (x+y))+d_3 (2 r_3 y-2 r_3 y r_1+r_2 (r_2 (2 r_3^2-1)+r_1^2)) \nonumber\\
&&-4 d_{22} r_1 r_2 (2 y r_3^2+u+2 (r_1^4+(-r_3+r_2 (u+x)+r_1 (-r_2^2+u+x)) r_1-u (x+y))) \nonumber\\
&&-4 d_{11} r_1 r_2 (-u+2 (u-2 r_1 r_3) y+2 r_1 (3 r_3 r_1^2-r_2 r_3 r_1+r_3+(3 r_1+r_2) (u-x)))),\\
A^{4'}_3 &=&\frac{{m_Z}^{7/2} N_0 \sqrt{r_3}}{r_1 r_2 L_1}(r_1-r_2)(d_1 T_b+d_3 T_c)
\end{eqnarray}
\end{widetext}

\subsection{Spin-triplet P-wave state: $|(c\bar{b})[^3P_J]\rangle$}

There are nine new Lorentz structures for $|(c\bar{b})[^3P_J]\rangle$, which are
\begin{widetext}
\begin{eqnarray}
B_1 &=&\frac{1}{m_Z}\varepsilon^J_{\alpha\alpha}k^{\nu},\;\;
B_2 =\frac{1}{m^3_Z}k^{\alpha}k^{\beta}\varepsilon^J_{\alpha\beta}k^{\nu}, \;\;
B_3 =\frac{1}{m^3_Z}k^{\alpha}q_2^{\beta}\varepsilon^J_{\alpha\beta}k^{\nu},\;\;
B_4 =\frac{1}{m^3_Z}q_2^{\alpha}q_2^{\beta}\varepsilon^J_{\alpha\beta}k^{\nu}, \nonumber\\
B_5 &=&\frac{i}{m^3_Z}\varepsilon(k,q_2,\beta,\alpha)\varepsilon^J_{\alpha\beta}k^{\nu}, \;\;
B_6 =\frac{i}{m^3_Z}\varepsilon(q_2,q_3,\beta,\alpha)\varepsilon^J_{\alpha\beta}k^{\nu}, \;\;
B_7 =\frac{i}{m^3_Z}\varepsilon(k,q_3,\beta,\alpha)\varepsilon^J_{\alpha\beta}k^{\nu}, \nonumber\\
B_8 &=&\frac{i}{m^3_Z}\varepsilon(k,q_2,q_3,\alpha)k^{\beta}\varepsilon^J_{\alpha\beta}k^{\nu},
B_9 =\frac{i}{m^3_Z}\varepsilon(k,q_2,q_3,\alpha)q_2^{\beta}\varepsilon^J_{\alpha\beta}k^{\nu},\nonumber
\end{eqnarray}
\end{widetext}

It is noted that $\varepsilon^{0,2}_{\alpha\beta}$ is the symmetric tensor and $\varepsilon^{1}_{\alpha\beta}$ is the anti-symmetric tensor, and the fact that $\varepsilon^{1}_{\alpha\alpha}=\varepsilon^{2}_{\alpha\alpha}=0$. As the terms involving the following coefficients have no contributions to the square of the amplitude. So practically we can safely set the coefficients before them to be zero:
\begin{equation}
\begin{array}{c}
A^i_j(|(c\bar{b})_{\bf 1}[^3P_0]\rangle)=0 \;\;\;\;\;\;\;\;{\rm for}\;i=(1-4),\\
j=(5,6,7,8,9)\\
\end{array}
\end{equation}
\begin{equation}
\begin{array}{c}
A^i_j(|(c\bar{b})_{\bf 1}[^3P_1]\rangle)=0 \;\;\;\;\;\;\;\;{\rm for}\;i=(1-4), \\
j=(1)\\
\end{array}
\end{equation}
\begin{equation}
\begin{array}{c}
A^i_j(|(c\bar{b})_{\bf 1}[^3P_2]\rangle)=0 \;\;\;\;\;\;\;\;{\rm for}\;i=(1-4),\\
j=(1,5,6,7,8,9) .
\end{array}
\end{equation}

The coefficients $A^{1}_j$, $A^{2}_j$,$A^{3'}_j$ and $A^{4'}_j$ that are the same for
all three $P$-waves:
\begin{widetext}
\begin{eqnarray}
A^1_1 &=&\frac{2 {L_1} {m_Z}^{5/2} }{{r_1} {r_2} \sqrt{{r_3}}}(2 {d_{21}} {r_2} {T_b} ({r_3}
-2 {r_1} x) {r_1}^2+2 {d_{32}} {r_2} {T_c} ({r_3}-2 {r_2} x) {r_1}^2+{d_4} (3 {r_1}+{r_2}) {r_3} {T_c}  \nonumber\\
&&\times({r_1} {r_3}+u)+{d_1} {T_b} ((3 {r_1}+{r_2}) {r_3} ({r_1} {r_3}+u)-2 {r_1} ({r_1}-{r_2}) x)
-{d_2} {r_3} {T_b} ({r_1} (3 {r_1}^2+{r_2}^2 \nonumber\\
&&+6 {r_2} {r_1})+(3 {r_1}+{r_2}) u-2 {r_3} y)+{d_3} {r_3} {T_c} (2 {r_1}^3+({r_3}^2+3 u) {r_1}+{r_2} u-2 {r_3} y)),\\
A^1_2 &=&-16 d_{11} {m_Z}^{5/2} r_1 \sqrt{r_3} T_b L_1,\\
A^1_4 &=&-8 {m_Z}^{5/2} {r_3}^{3/2} (d_{11} T_b-d_{22} T_b+(d_{31}+d_{41}) T_c) L_1,\\
A^1_5 &=&\frac{ {m_Z}^{5/2}\sqrt{r_3} L_1}{r_1 r_2}(d_2-d_3)(r_1-r_2),\\
A^1_6 &=&\frac{{m_Z}^{5/2}L_1}{2 r_1 r_2 r_3^{3/2}}(((3 d_1+3 d_2+d_3-d_4) r_1^2+2 (2 d_1-d_2+2 d_{21}
+d_3+2 d_{32}) r_2 r_1  \nonumber\\
&&+(d_1-d_2+5 d_3+d_4) r_2^2) r_3-8 r_1 r_2 (d_{21} r_1+d_{32} r_2) x),\\
A^1_7 &=&-\frac{{m_Z}^{5/2} \sqrt{r_3} L_1}{r_2}(d_1-d_2),\\
A^1_8 &=&-\frac{4{m_Z}^{5/2} L_1}{\sqrt{r_3}}(d_{21}+d_{22}+d_{32}),\\
A^1_9 &=&\frac{4{m_Z}^{5/2} L_1}{\sqrt{r_3}}(d_{22}-d_{31}),\\
A^2_1 &=&\frac{{L_2} {m_Z}^{5/2} }{2 {r_1} {r_2} \sqrt{{r_3}}}(2 {d_{21}} {r_2} ({r_3}-2 {r_1} x) {r_1}^2
+2 {d_{32}} {r_2} ({r_3}-2 {r_2} x) {r_1}^2+{d_4} (3 {r_1}+{r_2}) {r_3} ({r_1} {r_3}+u) \nonumber\\
&&+{d_1} (3 {r_1}^4+7 {r_2} {r_1}^3+(5 {r_2}^2+u-2 x) {r_1}^2+{r_2} ({r_2}^2-8 u+2 x) {r_1}-{r_2}^2 u)
+{d_2} (3 {r_3} {r_1}^3 \nonumber\\
&&+u {r_1}^2+{r_2} (6 u-{r_2} {r_3}) {r_1}+{r_2}^2 u-2 {r_3}^2 y)+{d_3} (3 {r_3} {r_1}^3+3 u {r_1}^2+2 {r_2} u {r_1} \nonumber\\
&&-{r_2}^2 r_3 {r_1}+3 {r_2}^2 u-2 {r_3}^2 y)),\\
A^2_2 &=&-4 d_{11} {m_Z}^{5/2} r_1 \sqrt{r_3} L_2,\\
A^2_4 &=&-2 {m_Z}^{5/2} ((3 d_{11}-d_{22}+d_{31}+d_{41}) r_1-(d_{11}-d_{22}+d_{31}-d_{41}) r_2) \sqrt{r_3} L_2,\\
A^2_5 &=&-\frac{4 {m_Z}^{5/2}\sqrt{r_3}L_2}{r_1 r_2}(d_2 T_b+d_3 T_c)(r_1-r_2),\\
A^2_6 &=&\frac{2 {m_Z}^{5/2} L_2}{r_1 r_2 r_3^{3/2}}(((d_1-d_2) r_1^2
+4 (d_2-d_{21}) r_2 r_1-(d_1-d_2) r_2^2) r_3 T_b+((d_3-d_4) r_1^2 \nonumber\\
&&-4 (d_3-d_{32}) r_2 r_1-(d_3-d_4) r_2^2) r_3 T_c+8 r_1 r_2 (d_{21} r_1 T_b-d_{32} r_2 T_c) x),\\
A^2_7 &=&-\frac{4{m_Z}^{5/2} \sqrt{r_3} T_b L_2}{r_2}(d_1-d_2),\\
A^2_8 &=&\frac{16 {m_Z}^{5/2} L_2}{\sqrt{r_3}}((d_{21}+d_{22}) T_b-d_{32} T_c),\\
A^2_9 &=&-\frac{16 {m_Z}^{5/2}L_2}{\sqrt{r_3}}(d_{22} T_b+d_{31} T_c),\\
A^{3'}_1 &=&\frac{1}{8 \sqrt{r_3} L_2}({m_Z}^{7/2} N_0 (\frac{2 d_4}{r_1 r_2}(r_1 (2 r_1^2
+5 r_2 r_1+r_2^2) r_3^2+(3 r_1+r_2) u r_3^2+(r_1-r_2) u^2)  \nonumber\\
&&-2 d_{21} (r_1 r_3+u) (2 r_1 x-r_3)+2 d_{32} (r_1 r_3+u) (r_3-2 r_2 x)
+\frac{d_3}{r_1 r_2}(2 (r_2^2+x-1) r_1^3 \nonumber\\
&&+4 r_2 (r_2^2+x-1) r_1^2+(2 (r_2^2+u+x-1) r_2^2-u+4 r_3^2 y+2 u y) r_1+r_2 (2 u r_2^2+u  \nonumber\\
&&+2 (5 r_2^2+3 u-3 x-2 y+1) r_1^3+2 r_2 (r_2^2-4 r_3^2 y-2 u y))
+\frac{d_1}{r_1 r_2}(6 r_1^5+14 r_2 r_1^4  \nonumber\\
&&+4 u-4 x-4 y+2) r_1^2+(2 (u-x-2 y+1) r_2^2+u (2 y-1)) r_1+r_2 u (1-2 y))  \nonumber\\
&&+\frac{d_2 }{r_1 r_2}(2 r_1^5+4 r_2 r_1^4+2 (r_2^2+u+x+y-1) r_1^3+2 r_2 (u+2 x+y-2) r_1^2 \nonumber\\
&&+(u+2 (r_2^2 (x-y-1)-u (x+y))) r_1-r_2 (2 y r_2^2+u-2 u x-2 u y)))),\\
A^{3'}_2 &=&-\frac{d_{11} {m_Z}^{7/2} N_0 \sqrt{r_3}}{L_2}(r_1 r_3+u),\\
A^{3'}_4 &=&-\frac{{m_Z}^{7/2} N_0 \sqrt{r_3}}{2 L_2}(d_{22}-d_{31}+d_{11} (2 r_1 r_3+2 x+2 y-1)
+2 (d_{41} (r_3^2+u) \nonumber\\
&&+d_{22} (r_1 r_3-x-y)+d_{31} (r_2 r_3+y))),\\
A^{3'}_5 &=&-\frac{{m_Z}^{7/2} N_0 \sqrt{r_3}}{r_1 r_2 L_2}(-d_1 T_b r_3^2+d_2 (r_1^2-r_2^2+r_3^2) T_b+d_3 (r_1-r_2) r_3 T_c),\\
A^{3'}_6 &=&\frac{1}{2 r_1 r_2 \sqrt{r_3} L_2}({m_Z}^{7/2} N_0 (4 d_{21} r_2 T_b x r_1^2
-2 d_{21} r_2 r_3 T_b r_1-2 d_4 r_3^2 T_c r_1-2 d_3 r_2 r_3 T_c r_1 \nonumber\\
&&+2 d_{32} r_2 r_3 T_c r_1-4 d_{32} r_2^2 T_c x r_1-d_3 r_3 T_c-2 d_4 r_3 T_c u+d_2 r_3 T_b (2 r_1 r_2-2 x-2 y+1)  \nonumber\\
&&+2 d_3 r_3 T_c y+d_1 r_3 T_b (4 x+2 y-1))),\\
A^{3'}_7 &=&-\frac{{m_Z}^{7/2} N_0 \sqrt{r_3}}{r_1 r_2 L_2}(d_3 r_1 (r_1-r_2) T_c-2 d_2 T_b (r_1 r_2+u)+d_1 T_b (r_1 r_3+2 u)),\\
A^{3'}_8 &=&\frac{4 {m_Z}^{7/2} N_0 \sqrt{r_3}}{L_2}(d_{11} T_b-d_{32} T_c),\\
A^{3'}_9 &=&-\frac{4 {m_Z}^{7/2} N_0 \sqrt{r_3}}{L_2}(d_{11} T_b+d_{31} T_c),\\
A^{4'}_1 &=&\frac{1}{2 \sqrt{r_3} L_1}({m_Z}^{7/2} N_0 (-\frac{2 d_4 T_c (r_1 r_3+u)}
{r_1 r_2}(r_1 (2 r_3^2+u)-r_2 u)+2 d_{21} T_b (r_1^2-r_2 r_1+u)  \nonumber\\
&&\times(2 r_1 x-r_3)-2 d_{32} T_c (r_1^2-r_2 r_1+u)(r_3-2 r_2 x)+\frac{d_3 T_c }{r_1 r_2}(2 r_2 r_1^4-2 (x-1) r_1^3 \nonumber\\
&&-2 r_2 (r_2^2-u+2 x-2) r_1^2+(2 (u-x+1) r_2^2+u-2 (2 r_3^2+u) y) r_1
+r_2 u (2 y-1)) \nonumber\\
&&-\frac{d_1 T_b }{r_1 r_2}(2 (x+2 y-1) r_1^3+4 r_2 (3 x+2 y-1) r_1^2+(2 (x+2 y-1) r_2^2+u (2 y-1)) r_1 \nonumber\\
&&+r_2 u (1-2 y))+\frac{d_2 T_b }{r_1 r_2}(2 r_2 r_1^4+2 (x+y-1) r_1^3 +2 r_2 (-r_2^2+u+2 x +3 y-2) r_1^2\nonumber\\
&&+(2 (u+x+3 y-1) r_2^2+u (2 x+2 y-1)) r_1+r_2 (2 y r_2^2+u-2 u x-2 u y)))),\\
A^{4'}_2 &=&\frac{4 d_{11} {m_Z}^{7/2} N_0 \sqrt{r_3} T_b }{L_1}(r_1^2-r_2 r_1+u),\\
A^{4'}_4 &=&\frac{2 {m_Z}^{7/2} N_0 \sqrt{r_3}}{L_1}((d_{11} T_b - d_{22} T_b )(2 x+2 y-1)
+T_c (d_{41} (r_1^2-r_2^2+r_3^2+2 u) \nonumber\\
&&+d_{31} (2 y-1))),\\
A^{4'}_5 &=&\frac{{m_Z}^{7/2} N_0 \sqrt{r_3}}{4 r_1 r_2 L_1}((d_2-d_1) r_3^2-(d_2-d_3) (r_1-r_2)^2),\\
A^{4'}_6 &=&\frac{1}{8 r_1 r_2 r_3^{3/2} L_1}({m_Z}^{7/2} N_0 (2 d_{21} r_1 (3 r_1-r_2)
r_2 (2 r_1 x-r_3)-2 d_{32} r_1 (r_1-3 r_2) r_2 (r_3-2 r_2 x) \nonumber\\
&&+r_3 (d_1 r_3+2 (d_4 (r_1^2+r_2^2+u)-d_1 (r_1^2-r_2 r_1+2 x+y)) r_3
+d_3 (2 r_2^3-6 r_1 r_2^2 \nonumber\\
&&+r_3-2 r_3 y)-d_2 (r_1 (2 r_1 (r_1-3 r_2)+1)+r_2-2 r_3 x-2 r_3 y)))),\\
A^{4'}_7 &=&\frac{{m_Z}^{7/2} N_0 \sqrt{r_3}}{4 r_1 r_2 L_1}((d_1-2 d_2+d_3) r_1 (r_1-r_2)+2 (d_1-d_2) u),\\
A^{4'}_8 &=&\frac{{m_Z}^{7/2} N_0}{\sqrt{r_3} L_1}((d_{21}+d_{22}+d_{32}) (r_1-r_2)+(-d_{11}+d_{21}+d_{22}) r_3),\\
A^{4'}_9 &=&\frac{{m_Z}^{7/2} N_0}{\sqrt{r_3} L_1}((d_{31}-d_{22}) (r_1-r_2)+(d_{11}-d_{22}) r_3),
\end{eqnarray}
\end{widetext}

Remaining non-zero coefficients for $|(c\bar{b})[^3P_0]\rangle$ and $|(c\bar{b})[^3P_2]\rangle$:
\begin{widetext}
\begin{eqnarray}
A^1_3 &=&8 L_1 {m_Z}^{5/2} \sqrt{r_3} (d_{11} (3 r_1+r_2) T_b-r_3 ((d_{21}+d_{22}) T_b+d_{32} T_c)),\\
A^2_3 &=&2 L_2 {m_Z}^{5/2} ((-d_{11}+d_{21}+d_{22}+d_{32}) r_2-(-5 d_{11}+d_{21}+d_{22}+d_{32}) r_1) \sqrt{r_3},\\
A^{3'}_3 &=&\frac{{m_Z}^{7/2} N_0 \sqrt{r_3} }{2 L_2}(2 r_1 r_3 d_{21}-2 x d_{21}
+d_{21}+d_{22}+d_{32}+2 d_{22} r_1 r_3-2 d_{32} r_2 r_3-2 d_{22} x  \nonumber\\
&&-2 (d_{21}+d_{22}+d_{32}) y+d_{11} (4 r_1 r_3+2 u+2 x+2 y-1)),\\
A^{4'}_3 &=&-\frac{2 {m_Z}^{7/2} N_0 \sqrt{r_3}}{L_1}(d_{32} T_c-(d_{21}
+d_{22}) T_b (2 x-1)-2 ((d_{21}+d_{22}) T_b+d_{32} T_c) y \nonumber\\
&&+d_{11} T_b (2 r_1 (r_1-r_2)+2 u+2 x+2 y-1)),
\end{eqnarray}
\end{widetext}

Remaining non-zero coefficients for $|(c\bar{b})[^3P_1]\rangle$:
\begin{widetext}
\begin{eqnarray}
A^1_3 &=&8 L_1 {m_Z}^{5/2} \sqrt{r_3} (d_{11} (r_1-r_2) T_b+(d_{21}+d_{22}) r_3 T_b+d_{32} r_3 T_c),\\
A^2_3 &=&-2 (d_{11}-d_{21}-d_{22}-d_{32}) L_2 {m_Z}^{5/2} (r_1-r_2) \sqrt{r_3},\\
A^{3'}_3 &=&-\frac{{m_Z}^{7/2} N_0 \sqrt{r_3}}{2 L_2 r_1 r_2}(r_1 r_2 (d_{21}+d_{22}
+d_{32}+2 ((d_{21}+d_{22}) r_1-d_{32} r_2) r_3-2 (d_{21}+d_{22}) x  \nonumber\\
&&-2 (d_{21}+d_{22}+d_{32}) y+d_{11} (-2 u+2 x+2 y-1))-d_1 r_3^2),\\
A^{4'}_3 &=&-\frac{2 {m_Z}^{7/2} N_0 \sqrt{r_3}}{L_1 r_1 r_2}(d_1 T_b r_3^2
+r_1 r_2 (-d_{32} T_c+(d_{21}+d_{22}) T_b (2 x-1) \nonumber\\
&&+d_{11} T_b (2 r_1 (r_1-r_2)+2 u-2 x-2 y+1)+2 ((d_{21}+d_{22}) T_b+d_{32} T_c) y)),
\end{eqnarray}
\end{widetext}

\section{Lorentz structure coefficients of $L^{\mu}_{rr^{\prime}}$}

All the non-zero coefficients of $L^{\mu}_{rr^{\prime}}$,
\begin{eqnarray}
C^{r-r^{\prime}}_1 &=&\frac{-m_Z N_1}{4 \sqrt{2} M}(1+4 T_d) p_2\cdot q_3\\
C^{r-r^{\prime}}_2 &=&\frac{-m_Z N_1}{4 \sqrt{2} M}(1+4 T_d) p_1\cdot q_3\\
C^{r-r^{\prime}}_3 &=&\frac{m_Z N_1}{4 \sqrt{2} M}(1+4 T_d) p_1\cdot p_2 \\
C^{r-r^{\prime}}_4 &=&\frac{-{m_Z}^3 N_1}{4 \sqrt{2} M}(1+4 T_d)\\
C^{-rr^{\prime}}_1 &=&\frac{m_Z N_1}{8 \sqrt{2} M}(\sqrt{2}-4(2-\sqrt{2}) T_d) p_2\cdot q_3\\
C^{-rr^{\prime}}_2 &=&\frac{m_Z N_1}{8 \sqrt{2} M}(\sqrt{2}-4(2-\sqrt{2}) T_d) p_1\cdot q_3\\
C^{-rr^{\prime}}_3 &=&\frac{m_Z N_1}{8 \sqrt{2} M}(-\sqrt{2}+4 (2-\sqrt{2}) T_d) ({p}_1\cdot {p}_2)\\
C^{-rr^{\prime}}_4 &=&\frac{{m_Z}^3 N_1}{8 M}(1+4 T_d-\sqrt{2}),
\end{eqnarray}
where $M=\frac{1}{\sqrt{p_1\cdot p_2}}$ and $T_d=\frac{1}{4}-{\sin ^2}{\theta _w}$. $N_1$ can be obtained by ensuring $k_1^\mu  = i{N_1}{\varepsilon ^{\mu \nu \rho \sigma }}{p_{1\nu }}{q_{3\rho }}{p_{2\sigma }}$ to satisfy $k^2_1=-1$.

\end{document}